\documentclass[usegraphicx,usenatbib,fleqn]{mn2e}

\usepackage{amsmath}
\usepackage{amssymb}
\usepackage{epsfig}
\usepackage{aas_macros}
\usepackage{txfonts}

\renewcommand\pi\piup

\def\Msun{{M_\odot}}
\def\rvir{r_{\rm vir}}
\def\Mt{M_{\rm t}}
\def\vlos{v_{\rm los}}
\def\Rin{R_{\rm in}}
\def\Rout{R_{\rm out}}
\def\rd{{\rm d}}

\def\rin{r_{\rm in}}
\def\rout{r_{\rm out}}
\def\sigmalos{\sigma_{\rm los}}
\def\sigmar{\sigma_{\rm r}}
\def\sigmat{\sigma_{\rm t}}
\def\kms{\mbox{km s$^{-1}$}}

\def\spose#1{\hbox to 0pt{#1\hss}}
\def\lta{\mathrel{\spose{\lower 3pt\hbox{$\sim$}} \raise 2.0pt\hbox{$<$}}}
\def\gta{\mathrel{\spose{\lower 3pt\hbox{$\sim$}} \raise 2.0pt\hbox{$>$}}}

\begin{document}

\voffset-.6in

\bibliographystyle{mn2e}

\title[The Masses of the Milky Way and Andromeda galaxies]{The Masses
  of the Milky Way and Andromeda Galaxies}

\author[Watkins, Evans \& An]{Laura~L.~Watkins$^1$, N.~Wyn~Evans$^1$, 
        Jin~H.~An$^{2,3,4}$
	\medskip
	\\$^1$Institute of Astronomy, University of Cambridge, Madingley Road, Cambridge, CB3 0HA, UK
\\$^2$ National Astronomical Observatories,
Chinese Academy of Sciences, A20 Datun Road, Chaoyang District,
Beijing 100012, PR China,
\\$^3$ Dark Cosmology Centre,
Niels Bohr Institutet, K{\o}benhavns Universitet,
Juliane Maries Vej 30, DK-2100 Copenhagen~\O, Denmark,
\\$^4$ Niels Bohr International Academy,
Niels Bohr Institutet, K{\o}benhavns Universitet,
Blegdamsvej 17, DK-2100 Copenhagen~\O, Denmark.}



\maketitle

\label{firstpage}

\begin{abstract}
  We present a family of robust tracer mass estimators to compute the
  enclosed mass of galaxy haloes from samples of discrete positional
  and kinematical data of tracers, such as halo stars, globular
  clusters and dwarf satellites. The data may be projected positions,
  distances, line of sight velocities or proper motions. The
  estimators all assume that the tracer population has a scale-free
  density and moves in a scale-free potential in the region of
  interest. The circumstances under which the boundary terms can be
  discarded and the estimator converges are derived. Forms of the
  estimator tailored for the Milky Way galaxy and for M31 are
  given. Monte Carlo simulations are used to quantify the uncertainty
  as a function of sample size.

  For the Milky Way galaxy, the satellite sample consists of 26
  galaxies with line-of-sight velocities. We find that the mass of the
  Milky Way within 300~kpc is $M_{300} = 0.9 \pm 0.3 \times 10^{12}
  \Msun$ assuming velocity isotropy. However, the mass estimate is
  sensitive to the assumed anisotropy and could plausibly lie between
  $0.7$ - $3.4$ $\times 10^{12} \Msun$, if anisotropies implied by
  simulations or by the observations are used. Incorporating the
  proper motions of 6 Milky Way satellites into the dataset, we find
  $M_{300} = 1.4 \pm 0.3 \times 10^{12} \Msun$. The range here if
  plausible anisotropies are used is still broader, from $1.2$ - $2.7
  \times 10^{12} \Msun$.  Note that our error bars only incorporate
  the statistical uncertainty. There are much greater uncertainties
  induced by velocity anisotropy and by selection of satellite
  members.

  For M31, there are 23 satellite galaxies with measured line-of-sight
  velocities, but only M33 and IC 10 have proper motions. We use the
  line of sight velocities and distances of the satellite galaxies to
  estimate the mass of M31 within 300~kpc as $M_{300} = 1.4 \pm 0.4
  \times 10^{12} \Msun$ assuming isotropy. There is only a modest
  dependence on anisotropy, with the mass varying between $1.3$ - $1.6
  \times 10^{12} \Msun$. Incorporating the proper motion dataset does
  not change the results significantly.  Given the uncertainties, we
  conclude that the satellite data by themselves yield no reliable
  insights into which of the two galaxies is actually the more
  massive.

  Leo I has long been known to dominate mass estimates for the Milky
  Way due to its substantial distance and line-of-sight velocity.  We
  find that And XII and And XIV similarly dominate the estimated mass
  of M31.  As such, we repeat the calculations without these galaxies,
  in case they are not bound -- although on the balance of the
  evidence, we favour their inclusion in mass calculations.
\end{abstract}



\section{Introduction}
\label{sect:introduction}

The structure and extent of dark matter haloes have important
implications for modern astrophysics, yet the determination of such
properties is a difficult task and the results are often conflicting.
A neat illustration is provided by the usage of Sagittarius Stream
data to constrain the shape of the Milky Way dark halo. This has told
us that the halo is nearly spherical \citep{2006ApJ...651..167F},
prolate \citep{2004ApJ...610L..97H}, oblate
\citep{2005ApJ...619..800J} or triaxial \citep{2009ApJ...703L..67L} in
nature! The Milky Way is the closest halo available for our study, the
availability of data has improved substantially in recent years, and
yet we are not able to determine its shape reliably.  

Similarly, we are unable to measure the masses of the Milky Way, or
its neighbour, the Andromeda Galaxy (M31) with any precision.  Despite
their proximity to us, their masses remain sketchily determined and
there is some controversy as to which halo is more massive.  Judged by
criteria such as the surface brightness of the stellar halo or the
numbers of globular clusters or the amplitude of the gas rotation
curve, M31 is seemingly the more massive. Judged by criteria such as
the velocities of the satellite galaxies and distant globulars or tidal
radii of the dwarf spheroidals, then the Milky Way is seemingly the
more massive. For example, \citet{2000ApJ...540L...9E} argued that the
M31 halo is roughly as massive as that of the Milky Way, with the
Milky Way marginally being the more massive of the two, while recent
studies have found evidence favouring both the Milky Way
\citep[e.g.][]{2000MNRAS.316..929E,2002MNRAS.337...34G} and M31
\citep[e.g.][]{2002ApJ...573..597K, 2009MNRAS.393.1265K} as the more
massive galaxy.

The masses of both haloes within a few tens of kiloparsecs are
reasonably well constrained by gas rotation curve data
\citep[e.g.][]{1988A&A...201...51R,1991ApJ...372...54B}. However,
these data only sample the inner parts of the haloes.  In order to
probe further out, we must turn to the kinematics of the satellite
populations. Such tracers are a valuable tool for studying the dark
matter haloes as their orbits contain important information about
their host potential.  Distance, radial velocity and proper motion
data can be used to constrain halo extent, mass and velocity
anisotropy \citep[see
e.g.][]{1987ApJ...320..493L,1996ApJ...457..228K,1999MNRAS.310..645W}.

The uncertainties in the mass estimates for the Milky Way and M31 are
largely due to the fact that there is seldom proper motion data
available to complement distance and radial velocity information.
With only one velocity component to work with, the eccentricities of
the orbits are poorly constrained. Statistical methods must be applied
to determine masses and these methods suffer greatly from the small
sample sizes available, even with the recent burst of satellite
discoveries associated with both galaxies.

The projected mass estimator was introduced by
\citet{1981ApJ...244..805B}. They assumed that only projected distance
and line-of-sight velocity information was available. The estimator is
also contained in the study of \citet{1981MNRAS.195.1037W} on
scale-free ensembles of binary galaxies. The analysis was extended by
\citet{1985ApJ...298....8H} and further modified by
\citet{2003ApJ...583..752E} to consider the case of tracer
populations.  These previous studies successfully used the mass
estimator to weigh M31. However, in its present form, the mass
estimator is ill-suited for application to the Milky Way and such a
study has not yet been attempted.

Here, we develop alternative forms of the estimator, and analyse the
conditions under which they are valid.  In addition, the census of
satellites around M31 has increased significantly
\citep{2004ApJ...612L.121Z, 2007ApJ...659L..21Z, 2006MNRAS.371.1983M,
  2007ApJ...670L...9M, 2007ApJ...671.1591I, 2008ApJ...676L..17I,
  2008ApJ...688.1009M} since the last studies of this type were
attempted and so we have more data at our disposal.  Hence, we apply
our estimator to M31 with these new data.


\section{Mass estimators}
\label{sect:massestimators}

The projected mass estimator \citep{1981ApJ...244..805B} takes the
form
\begin{equation}\label{eq:BTest}
  M = \frac{C}{G} \left< \vlos^2 R \right> = 
\frac{C}{G}\frac{1}{N}\sum_{i=1}^N v_{{\rm los},i}^2 R_i
\end{equation}
for a set of $N$ tracers objects (e.g planetary nebulae, stars,
globular clusters, dwarf spheroidal galaxy satellites) with
line-of-sight velocities $\vlos$ and projected distances $R$. Here,
$G$ is the gravitational constant and $C$ is a constant determined by
the host potential and the eccentricity of the orbits. They found that
$C = 16/ \pi$ for test particles with an isotropic velocity
distribution orbiting a point mass and $C = 32/ \pi$ for test
particles moving on radial orbits.

This analysis was extended by \citet{1985ApJ...298....8H} to consider
the case in which tracers may track the total mass (e.g. in galaxy
groups).  They found that $C = 32/\pi$ for particles with an isotropic
velocity distribution and $C = 64/ \pi$ for particles on radial
orbits.  A key assumption in this work is that the members/tracers
track the mass of the group/host.  This is not true for all tracer
populations, particularly for those tracers which are commonly used to
estimate the masses of ellipticals or the haloes of spiral galaxies.

\subsection{Tracer Mass Estimator}
\label{sect:tracerestimator}

Here, we give a formal derivation of our tracer estimators, so as to
clarify the conditions under which they converge to the enclosed
mass. Readers primarily interested in applications, and willing to
take convergence on trust, should skip straight to the estimators
themselves, namely eqns~(\ref{eq:firstcase}), (\ref{eq:Andcase}),
(\ref{eq:PMcase}) and (\ref{eq:lastcase}).  We give formulae for the
various cases in which true distances or projected distances, and
line-of-sight velocities, or radial velocities or proper motions, are
known for the tracers. The estimators are both simple and flexible.

Let us begin by supposing that the observations are discrete positions
$r$ and radial velocities $v_r$ of $N$ members of a tracer
population. Here, $r$ is measured from the centre of the host galaxy,
whilst $v_r = {\dot r}$ is the radial velocity. We propose to combine
the positional and kinematic data to give the enclosed mass $M$ in the
form
\begin{equation}
	M  = \frac{C}{G} \left< v_r^2 r^{\lambda} \right> =
\frac{C}{G} \frac{1}{N} \sum_{i=1}^N v_{r,i}^2 r_i^\lambda.
\end{equation}
Here, unlike equation~(\ref{eq:BTest}),
the constant $C$ is not necessarily dimensionless.
Notice that {\it a priori} we do not know the best choice for
$\lambda$. This will emerge from our analysis.

If $f$ is the phase space distribution function of the tracers and
$\sigmar$ the radial velocity dispersion, we see that under the
assumption of spherical symmetry:
\begin{equation}
\langle v_r^2 r^\lambda\rangle
=\frac1\Mt
\int\!\rd^3\!\bmath r\,\rd^3\!\bmath v\,
f v_r^2 r^\lambda
=\frac{4\pi}{\Mt}
\int\rho\sigmar^2r^{\lambda+2}\,\rd r
\end{equation}
where $\Mt$ is the mass in the tracers
\begin{equation}
\Mt=
4\pi\int r^2\rho\,\rd r.
\end{equation}
Now, let us assume that the tracer population is spherically symmetric and
has a number density which falls off like a power-law
\begin{equation}
\rho(r)\propto r^{-\gamma}\,;\qquad
\frac{\rd\log\rho}{\rd\log r}=-\gamma
\label{eq:cuspdens}
\end{equation}
at least within the radius interval [$\rin,\rout$] where the data lie.
Then, the estimator reduces to
%
\begin{equation}
\label{eq:tra}
\langle v_r^2 r^\lambda\rangle
=\frac1{\mathcal M}
\int_{\rin}^{\rout}r^{\lambda-\gamma+2}\sigmar^2\,\rd r
\,;\qquad
\mathcal M=\begin{cases}\
\displaystyle{\frac{\rout^{3-\gamma}-\rin^{3-\gamma}}{3-\gamma}}
&(\gamma\ne3)\medskip\\\
\log\,\biggl(\dfrac{\rout}{\rin}\biggr)&(\gamma=3)
\end{cases},
\end{equation}
%
where $\log x$ is the natural logarithm.
Once the behaviour of $\sigmar^2$ is found, we may relate this
estimator to the dynamical halo mass $M(r)$.  This can be achieved
through solving the Jeans equation, which reads:
\begin{equation}
{1\over \rho} {\mathrm{d} (\rho \sigmar^2) \over \mathrm{d}r}
+ {2\beta \sigmar^2
\over r} = -{GM(r)\over r^2}.
\label{eq:jeans}
\end{equation}
Here, we have introduced $\beta = 1- {\sigmat^2}/{\sigmar^2}$, the
Binney anisotropy parameter, in which $\sigmat$ is the tangential
velocity dispersion. Now, $\beta \to \infty$ corresponds to a circular
orbit model, $\beta = 1$ corresponds to purely radial orbits and
$\beta = 0$ is the isotropic case.
We note that the Jeans equation (\ref{eq:jeans}) in a spherical system
can be put into the form
\begin{equation}
  Q\rho\sigmar^2
  =-\int\!Q\,\rho\,\frac{GM(r)}{r^2}\,\rd r\,;\qquad
  \log Q=\int\frac{2\beta}r\,\rd r.
\end{equation}
If $\beta$ is independent of $r$, this simplifies to be $Q=r^{2\beta}$.

To proceed further, the underlying gravity field is assumed to be
scale-free at least in the interval [$\rin,\rout$], that is, the
relative potential up to a constant is given by
\begin{equation}
\psi(r)=\begin{cases}\
\displaystyle{		
\frac{v_0^2}{\alpha} \left( \frac{a}{r} \right)^{\alpha} }
&(\alpha \ne 0)
\medskip\\\
v_0^2\, \log\left(\dfrac{a}{r}\right) & (\alpha = 0)
\end{cases}
\label{eq:gravfield}
\end{equation}
with $-1 \le \alpha \le 1$.\footnote{$\alpha=-1$ corresponds to the
  gravitational field that pulls with an equal magnitude force
  regardless of radius, which is formally generated by a halo density
  falling off as $r^{-1}$. Provided we regard the scale-free potential
  as an approximation valid over a limited range and not extending to
  spatial infinity, we can permit $\alpha \ge -2$, since $\alpha =-2$
  corresponds to the harmonic potential generated by a homogeneous
  sphere.} Here, $a$ is a fiducial radius, which should lie in the
region for which the power-law approximation for the relative
potential is valid (i.e., $\rin\le a\le\rout$) and $v_0$ is the
circular speed at that radius $a$.  When $\alpha=1$, this corresponds
to the case in which the test particles are orbiting a point-mass;
when $\alpha =0$, the satellites are moving in a large-scale mass
distribution with a flat rotation curve; when $\alpha = \gamma - 2$,
the satellites track the total gravitating mass. We remark that our
model of a scale-free tracer population of satellites in a scale-free
potential has previously been used to study the mass of the Milky Way
by ~\citet{1992MNRAS.255..105K}, although using the standard technique
of maximum likelihood for parameter estimation.

The scale-free assumption is also equivalent to proposing the halo mass
profile to be
\begin{equation}
\frac{M(r)}{M(a)}=\left(\frac{r}{a}\right)^{1-\alpha},
\end{equation}
and the local mass density $\propto r^{-(\alpha+2)}$.  Consequently,
if the power-law behaviour were allowed to be extended to infinity,
the total mass of the dark halo would necessarily be infinite unless
$\alpha=1$.  (However, if the halo density were to fall off faster
than $r^{-3}$ and so the total gravitating mass is finite, the leading
term for the potential would be Keplerian. That is to say, for the
case of a finite total mass halo, the gravity field experienced by the
tracers may be approximated to be that of a point mass, given that
$\rin$ is chosen to be sufficiently large so that the gravitating mass
inside the sphere of $\rin$ dominates the mass within the shell region
populated by the tracers.)

Combining this with the constant-anisotropy assumption, the Jeans
equation integrated between $r$ and $\rout$ then reduces to
\begin{equation}\label{eq:jeansint}
r^{2\beta-\gamma}\sigmar^2(r)
-\rout^{2\beta-\gamma}\sigmar^2(\rout)
=\frac{GM(a)}{a^{1-\alpha}}
\int_r^{\rout}\tilde r^{2\beta-\gamma-\alpha-1}\rd\tilde r.
\end{equation}
provided that all our assumptions remain valid in the radius interval
$[\rin,\rout]$ and $r,a\in[\rin,\rout]$.

Now, our goal is to find the total halo mass. In reality, the observed
tracers are only populated up to a finite outer radius, and so, any
mass distribution outside of that radius does not affect our
observations in a strictly spherical system (Newton's theorem). We
therefore extend the power-law potential assumption only up to the
finite outer radius (here $\rout$), and set $a=\rout$. In other
words, the halo mass that we are interested in is that contained
within the outer radius, $M=M(\rout)$. With $a=\rout$,
solving equation (\ref{eq:jeansint})
for $\sigmar^2(r)$ results in (here $s\equiv r/\rout$)
\begin{equation}\label{eq:sol}
\sigmar^2=\begin{cases}\
\dfrac{\sigmar^2(\rout)-\hat v_0^2}{s^{2\beta-\gamma}}
+\dfrac{\hat v_0^2}{s^\alpha}
&(\alpha+\gamma-2\beta\ne0)\medskip\\\
\displaystyle{\frac{\sigmar^2(\rout)-v_0^2\,\log s}{s^\alpha}}
&(\alpha=2\beta-\gamma)
\end{cases}
\end{equation}
where $v_0^2=GM/\rout$ is the circular speed at $\rout$ whilst
$\hat v_0^2\equiv v_0^2/(\alpha+\gamma-2\beta)$.

Then, substituting the result of equation (\ref{eq:sol})
into equation (\ref{eq:tra}) and
explicitly performing the integration yields
(ignoring particular parameter combinations that involve the logarithm)
\begin{multline}\label{eq:vr}
\frac{\langle v_r^2 r^\lambda\rangle}{(3-\gamma)\rout^\lambda}
=\frac{v_0^2}{(\lambda-\alpha+3-\gamma)(\alpha+\gamma-2\beta)}
\frac{1-u^{\lambda-\alpha+3-\gamma}}{1-u^{3-\gamma}}
\\+\frac1{\lambda-2\beta+3}
\left[\sigmar^2(\rout)-
\frac{v_0^2}{\alpha+\gamma-2\beta}\right]
\frac{1-u^{\lambda-2\beta+3}}{1-u^{3-\gamma}}
\end{multline}
where $u\equiv\rin/\rout$. Notice now that the choice of
$\lambda=\alpha$ makes the $u$-dependence of the first term in the
right-hand side drop out. In fact, this could also have been deduced
on dimensional grounds by requiring that our estimator is not
dominated by datapoints at small radii or large radii.

The last terms in equation (\ref{eq:vr}) basically constitute the
surface `pressure' support terms in the Jeans equation, which we wish
to minimize as $u\rightarrow0$. Here, we limit ourselves to the case
that $\lambda=\alpha$, when the corresponding leading term is
\begin{equation}
\frac{1-u^{\alpha-2\beta+3}}{1-u^{3-\gamma}}
\sim\begin{cases}
1&2\beta-\alpha,\gamma<3
\\-u^{-(2\beta-\alpha-3)}
&\gamma<3<2\beta-\alpha
\\-u^{\gamma-3}
&2\beta-\alpha<3<\gamma
\\u^{\alpha+\gamma-2\beta}
&3<2\beta-\alpha,\gamma
\end{cases}.
\end{equation}
In other words, provided that $\gamma>3$ and $\gamma>2\beta-\alpha$,
the pressure term vanishes as $u\rightarrow0$, and we obtain the
scale-free Jeans solutions of \citet{1997MNRAS.286..315E}.  In fact,
since $\beta\le1$ and $-1\le\alpha\le1$, we find that
$2\beta-\alpha\le 3$ and thus the second condition here is essentially
redundant. Consequently, provided that $\gamma>3$, that is the tracer
density falls off more quickly than $r^{-3}$, we find the estimator to
be
\begin{equation}\label{eq:msest}
\langle v_r^2 r^\alpha\rangle\simeq
\frac{\rout^\alpha}{\alpha+\gamma-2\beta}
\frac{GM}{\rout}+\mathcal R
\end{equation}
where the remainder $\mathcal R\rightarrow0$ vanishes as
$\rin/\rout\rightarrow0$ (here, $\rin$ and $\rout$ are the inner and
outer radius of the tracer population).

Alternatively, if $\gamma<3$ and $2\beta-\alpha<3$,
the remainder term tends to a constant as $u\rightarrow0$.
In a perfectly scale-free halo traced by again strictly scale-free
populations, this constant must be zero. This is because, for such
a system, $\sigmar^2$ should also be scale-free. Yet equation (\ref{eq:sol})
implies that this is possible only if $\sigmar^2(\rout)=\hat v_0^2$.
Subsequently this also indicates that the coefficient for the remainder in
equation (\ref{eq:vr}) vanishes too. Even after relaxing the everywhere
strict power-law behaviour, we would expect that
$\sigmar^2\sim\hat v_0^2$ and consequently that
$|\sigmar^2-\hat v_0^2|\ll\hat v_0^2$, provided that
$2\beta-\alpha<\gamma$, which is required to ensure $\hat v_0^2>0$.
That is to say, we expect that $\hat v_0^2\rout^\alpha\gg\mathcal R$
as $u\rightarrow0$
in equation (\ref{eq:msest}) for $2\beta-\alpha<\gamma<3$,
which is sufficient for justifying the applicability of our
mass estimator. 

In other words, we have obtained a very simple result
\begin{equation}
M = \frac{C}{G} \left< v_r^2 r^{\alpha} \right>,
\quad\quad
C = \left( \alpha + \gamma -2 \beta \right) \rout^{1- \alpha},
\label{eq:firstcase}
\end{equation}
provided that $C>0$ (the simple interpolative argument indicates that
this is still valid for $\gamma=3$).  This corresponds to the case in
which the tracers have known radial velocity components $v_r$ resolved
with respect to the centre of the galaxy, as well as actual distances
$r$. For satellites of the Milky Way, the line of sight velocity
$\vlos$ is measured, and corrected to the Galactic rest frame.  Now,
$v_r$ may be calculated from $\vlos$ only if proper motion data
exists. Alternatively, a statistical correction can be applied to
estimate $v_r$ from $\vlos$
\begin{equation}\label{eq:v2sc}
\langle v_r^2\rangle=\frac{\langle\vlos^2\rangle}{1-\beta\sin^2\varphi}
\end{equation}
where $\varphi$ is the angle between the unit vector from the Galactic
Centre to the satellite and the unit vector from the Sun to the
satellite.

Note too that in the important isothermal case ($\alpha =0$), the
galaxy rotation curve is flat with amplitude $v_0$. Then, for members
of a population with density falling like $\rho \sim r^{-3}$, such as
the Galactic globular clusters, eqn~(\ref{eq:firstcase}) reduces to
\begin{equation}
v_0^2 = (3 - 2\beta) \langle v_r^2 \rangle.
\end{equation}
This is a generalization of the estimator of
\citet{1981Obs...101..200L} to the case of anisotropy. When the
population is isotropic ($\beta = 0$), it reduces to the appealing
simple statement that the circular speed is the rms velocity of the
tracers multiplied by $\sqrt3\approx1.732$.

Even if three dimensional distance $r$ is replaced by projected
distance $R$ or $v_r$ by some other projections of the velocity,
the basic scaling result of eqn~(\ref{eq:firstcase}) remains
valid. Different projections simply result in distinct constants $C$,
as we now show.

\begin{figure*}
\begin{center}
	\includegraphics[width=0.9\textwidth]{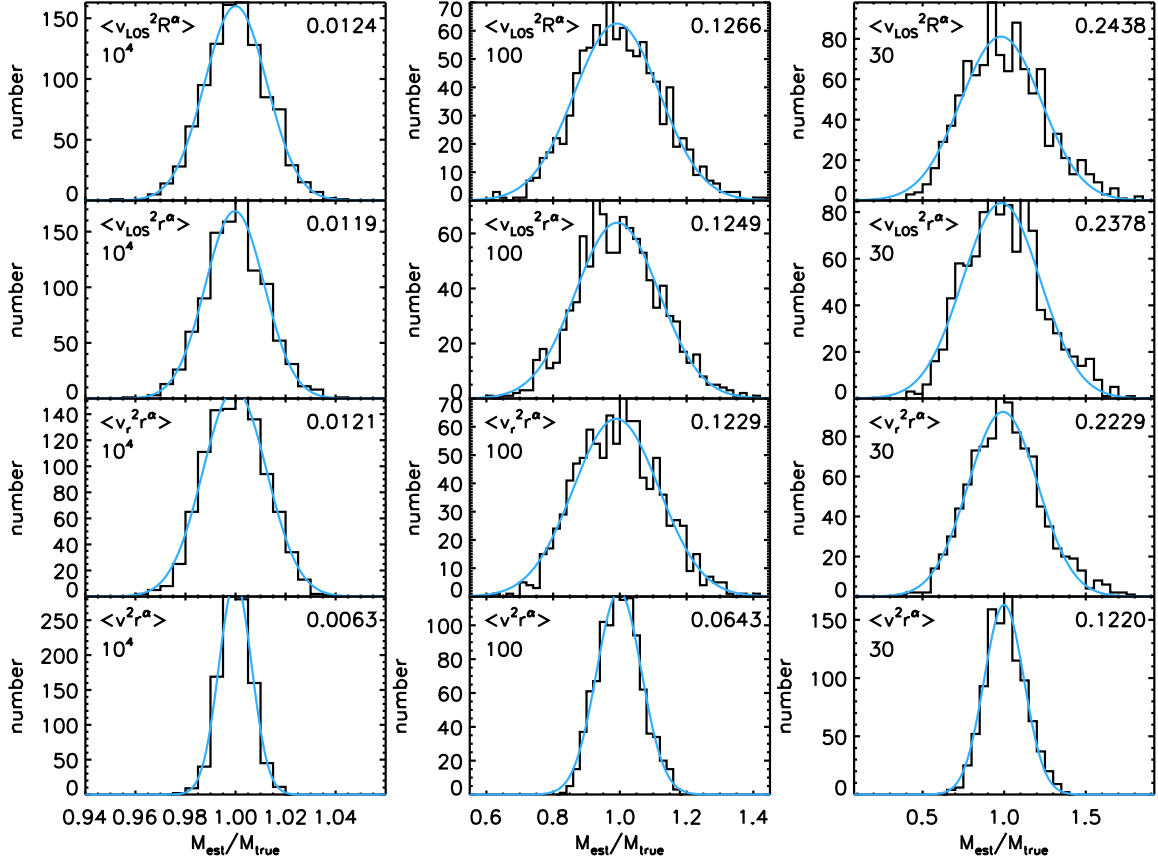}
	\caption{Distribution of mass estimate as a fraction of the
          true mass for 1000 Monte Carlo realisations, assuming that
          parameters $\alpha$, $\beta$ and $\gamma$ are known
          exactly. Left: $N=10,000$. Middle: $N=100$.  Right: $N=30$.
          The number of satellites in the simulation and the form of
          the estimator used to recover the mass is shown in the top
          left corner of each panel.  A best-fit Gaussian is plotted
          for each distribution and the standard deviation of the
          distribution is shown in the top right corner of each panel.
          On average, the tracer mass estimator recovers the true mass
          of the host. [The cases shown correspond to $\alpha = 0.55$,
          $\beta = 0.0$ and $\gamma = 2.7$].}
	\label{fig:masshists}
\end{center}
\end{figure*}

\subsection{A Family of Estimators}
\label{sect:modifyestimators}

Now, suppose that we have actual distances $r$ from the centre of the
host galaxy, but only projected or line of sight velocities
$\vlos$. This is the case for many of M31's satellite galaxies, for
which distances have been measured by using the tip of the red giant
branch method \citep[see e.g.,][]{2005MNRAS.356..979M} and for which
projected velocities are known from spectroscopy. The calculation
proceeds by considering
\begin{equation}
\langle\vlos^2r^\alpha\rangle
=\frac1\Mt
\int\!\rd^3\!\bmath r\,\rd^3\!\bmath v\,f\vlos^2r^\alpha
=\frac{2\pi}\Mt
\int_{\rin}^{\rout}\!\rd r\!\int_0^\pi\!\rd\theta\,
\rho\sigmalos^2r^{\alpha+2}\sin\theta
\end{equation}
We now need the relationship between line-of-sight velocity dispersion
$\sigmalos$ and the radial velocity dispersion $\sigmar$, namely
\begin{equation}
        \sigmalos^2 = \sigmar^2 \left( 1- \beta \sin^2 \varphi \right),
\end{equation}
which is similar to equation (\ref{eq:v2sc}) but here the angle
$\varphi$ is the angle between the line of sight and the position
vector of the satellite with respect to the centre of the \emph{host
  galaxy} \citep[see e.g.,][ Section 4.2]{1987gady.book.....B}.  If
the polar $z$-axis of the coordinate system is chosen such that the
sun (that is, the observer) lies on the negative $z$-axis (i.e.,
$\theta=\pi$) at a distance $d$ from the centre of the host galaxy, we
find that
\begin{equation}
\sin^2\!\varphi=\frac{\sin^2\!\theta}
{1+2\tfrac rd\cos\theta+\bigl(\tfrac rd\bigr)^2}.
\end{equation}
However, for most external galaxies, it is reasonable to assume
$d\gg\rout$, and therefore, we can safely approximate\footnote{On the
  other hand, for the satellites of the Milky Way, it is often assumed
  that $d\ll\rin$, which leads to $\sin\varphi\approx0$ and
  consequently $\langle\vlos^2r^\alpha\rangle \approx \langle
  v_r^2r^\alpha\rangle$.}  that $\sin^2\varphi\approx\sin^2\theta$.
Then,
\begin{equation}
\langle\vlos^2r^\alpha\rangle=
\langle v_r^2r^\alpha\rangle
\int_0^{\pi/2}\rd\theta\,\sin\theta\left(1-\beta\sin^2\theta\right),
\end{equation}
and thus we find that
\begin{equation}
M = \frac{C}{G} \left<\vlos^2 r^{\alpha} \right>,
\quad\quad
C = \frac{3\left( \alpha + \gamma -2 \beta \right)}
{3-2 \beta} \rout^{1- \alpha}.
\label{eq:Andcase}
\end{equation}

Next, we consider the case in which we have full velocity information
for the satellites, i.e., both radial velocities and proper
motions. For example, this is the case for a subset of the satellites
of the Milky Way~\citep[see e.g.,][]{2002AJ....124.3198P}. In this
case, we can utilize
$\sigma^2=\sigmar^2+\sigmat^2=(3-2\beta)\sigmar^2$, and therefore
the estimator becomes
\begin{equation}
M = \frac{C}{G} \left< v^2 r^{\alpha} \right>,\quad\quad
C = \frac{\alpha + \gamma -2 \beta}{3 - 2 \beta} \rout^{1- \alpha}.
\label{eq:PMcase}
\end{equation}

Finally, we can assume a worst-case scenario in which the only data
available are projected distances $R$ and line-of-sight velocities
$\vlos$ for the tracers. Outside of the galaxies of the Local Group,
this is the usual state of affairs. So, this would be the form of the
estimator to find the dark matter mass of nearby giant ellipticals like
M87 from positions and velocities of the globular clusters.
The estimator is derived following the same procedure
with $R=r\sin\theta$, which results in the relation
\begin{equation}
\langle\vlos^2R^\alpha\rangle=
\langle v_r^2r^\alpha\rangle
\int_0^{\pi/2}\rd\theta\,\sin^{\alpha+1}\!\theta
\left(1-\beta\sin^2\theta\right).
\end{equation}
Consequently, the corresponding estimator is found to be~\footnote{
  The result is valid provided that the integral is limited to
  spherical shells. However, given the lack of depth information, it
  might seem more logical to perform the integration over cylindrical
  shells. Unfortunately, the result is more complicated, as it
  involves the integrals of incomplete beta functions.}
\begin{equation}
M = \frac{C}{G} \left<\vlos^2 R^{\alpha} \right>, \quad\quad
C = \frac{\left( \alpha + \gamma -2 \beta \right)}
{I_{\alpha , \beta}} \rout^{1- \alpha}
\label{eq:lastcase}
\end{equation}
where
\begin{equation}
I_{\alpha,\beta}
=\frac{\pi^{1/2}\Gamma(\tfrac\alpha2+1)}{4\Gamma(\tfrac\alpha2+\tfrac52)}
\left[\alpha+3-\beta(\alpha+2)\right]
\end{equation}
and $\Gamma(x)$ is the gamma function.  This case is related to work
by Bahcall \& Tremaine (1981).  So, for example, in the Keplerian case
($\alpha =1$), a distribution of test particles with $\gamma =3$ gives
\begin{equation}
C = {32 \over \pi}{ 2- \beta \over 4 -3\beta}
\end{equation}
When $\beta =0$, this implies that $C = 16/\pi$; whilst when $\beta
=1$, $C = 32/ \pi$. 

Some of these estimators are implicit in other work. In particular,
some are equivalent to those introduced by
\citet{1981MNRAS.195.1037W}, who had a different focus on the dynamics
of binary galaxies but who made the same scale-free assumptions to
obtain robust mass estimators. Very recently, An \& Evans (2010, in
preparation) found a related family of estimators that are independent
of parameters derived from the tracer density (like $\gamma$).


\section{Checks with Monte Carlo Simulations}
\label{sect:montecarlo}

In order to verify the correctness of our mass estimators, we generate
synthetic data-sets of anisotropic spherical tracer populations.
Distances $r$ are selected in $\left[ \rin, \rout \right]$ assuming
the power-law density profile in eqn~(\ref{eq:cuspdens}).  Projection
directions are determined by the position angles: $\cos \theta$ is
generated uniformly in $\left[ -1, 1 \right]$ and $\phi$ is generated
uniformly in $\left[ 0,2 \pi \right]$. If $R$ lies outside of the
allowed range, the projection direction is regenerated until $R$ is
within $\left[ \Rin, \Rout \right]$.

The phase-space distribution functions that give rise to such density
profiles are given in \citet{1997MNRAS.286..315E}.  Tracer velocities
are picked from the distributions
\begin{equation}
f(v) \propto \begin{cases}\
	v^{2-2\beta} \left| \psi(r) - \tfrac12 v^2 \right|^{ \left[
            2 \gamma -3 \alpha -2 \beta \left( 2 - \alpha \right)
          \right] / \left( 2 \alpha \right) } & (\alpha\ne0)
\medskip\\\
	v^{2-2\beta} \exp \biggl( -\dfrac{v^2}{2\sigma^2} \biggr) & (\alpha=0)
\end{cases}.
\end{equation}
For $\alpha > 0$, the maximum velocity at any position is
$\sqrt{2\psi(r)}$; for $\alpha \le 0$, the velocities
can become arbitrarily large.  Following \citet{1987gady.book.....B},
we introduce spherical polar coordinates in velocity space $\left( v,
  \xi, \eta \right)$ so that the velocities resolved in spherical
polar coordinates with respect to the centre are then
\begin{equation}
	\begin{array}{lll}
	v_r = v \cos \eta &
	v_{\theta} = v \sin \eta \cos \xi &
	v_{\phi} = v \sin \eta \sin \xi
	\end{array}
\end{equation}
To generate velocities with the correct anisotropy, $\xi$ is generated
uniformly in $\left[ 0,2 \pi \right]$ and $\eta$ is picked in $\left[
  0, \pi \right]$ from the distribution
\begin{equation}
	F(\eta) \propto \left| \sin \eta \right| ^{1-2 \beta}
\end{equation}
where $\beta$ is the Binney anisotropy parameter.  Finally, the
line-of-sight velocities are calculated and used in the tracer mass
estimator.

Figure \ref{fig:masshists} shows the distribution of mass estimates as
fractions of the true mass for 1000 realisations, assuming that
parameters $\alpha$, $\beta$ and $\gamma$ are known exactly; the left
panels show simulations with 10,000 tracers, the middle panels for 100
tracers and the right panels for 30 tracers. The panels use the
different forms of the estimator given in eqns~(\ref{eq:firstcase}),
(\ref{eq:Andcase}), (\ref{eq:PMcase}) and (\ref{eq:lastcase})
respectively.  A Gaussian with the same standard deviation as each
distribution is also plotted for each panel. The standard deviation is
included in the top-right corner of each plot and gives an estimate of
the error in each case.

We see that our mass estimators are unbiased -- that is, on average,
the true mass is recovered in all cases. The benefit of using three
dimensional distances $r$ instead of projected distances $R$ is
modest, as is the improvement gained by using $v_r$ in place of
$\vlos$. However, if proper motion data are available, then using $v$
instead of $v_r$ gives a more accurate mass estimate.

So far, we have assumed that we know $\alpha$, $\beta$ and $\gamma$
exactly, which is, of course, not the case. Our estimates for
$\alpha$, $\beta$ and $\gamma$ have errors associated with them, not
least because the notion of a scale-free density profile in a
scale-free potential is an idealization. As these parameters enter the
estimator through the prefactor $C$, it is straightforward to obtain
the additional uncertainty in the final answer using propagation of
errors. As we will show in the next section $\alpha$ and $\gamma$ are
constrained either by cosmological arguments or by the data. The
right-most column in Figure \ref{fig:masshists} (a host with 30
satellites) is the most applicable to our data-sets at present as the
Milky Way has 26 satellites and M31 23 satellites with a recorded
line-of-sight velocity. The error on the mass estimate obtained in
this case is $\sim 25\%$. This is much larger than that the effects of
errors on $\alpha$ and $\gamma$ and so the latter will be ignored for
the rest of the discussion.

However, the case of the velocity anisotropy $\beta$ is different as
it is poorly constrained, with theory and data pointing in rather
different directions. Changes in $\beta$ can therefore make a
substantial difference to the mass estimate.

Note that these simulations yield no insight into systematic errors,
because the mock data are drawn from the same distribution functions
used to derive the form of the mass estimators. This is a concern as
there are a number of causes of systematic error -- for example, dark
halos may not be spherical, or infall may continue to the present day
so that the observed satellites may not necessarily be
virialized. Deason et al. (2010, in preparation) have tested the
estimators derived in this paper, as well as a number of other
commonly used mass estimators, against simulations. Specifically, they
extracted samples of Milky Way-like galaxies and their satellites from
the {\it Galaxies Intergalactic Medium Interaction
  Calculation}~\citep{2009MNRAS.399.1773C}, a recent high resolution
hydrodynamical simulation of a large volume of the Universe. They find
that the estimators in this paper significantly out-perform the
projected mass estimator of \citet{1981ApJ...244..805B} and the tracer
mass estimator of \citet{2003ApJ...583..752E}.


\begin{table}
\begin{center}
  \caption{Data table for the satellites of the Milky Way.  Listed are
    Galactic coordinates ($l$, $b$) in degrees, Galactocentric
    distance $r$ in kpc and corrected line-of-sight velocity in \kms.}
	\label{table:mwsats}
\begin{tabular}{lrrrrc}
\hline
\hline
Name & \multicolumn{1}{c}{$l$} & \multicolumn{1}{c}{$b$} & \multicolumn{1}{c}{$r$} & \multicolumn{1}{c}{$\vlos$} & \multicolumn{1}{c}{Source} \\
 & \multicolumn{1}{c}{(deg)} & \multicolumn{1}{c}{(deg)} & \multicolumn{1}{c}{(kpc)} & \multicolumn{1}{c}{(\kms)} & \\
\hline
         Bootes I & 358.1 &  69.6 &   57 &  106.6 & 1,2 \\
        Bootes II & 353.8 &  68.8 &   43 & -115.6 & 3,4 \\
 Canes Venatici I &  74.3 &  79.8 &  219 &   76.8 & 5,6 \\
Canes Venatici II & 113.6 &  82.7 &  150 &  -96.1 & 6,7 \\
           Carina & 260.1 & -22.2 &  102 &   14.3 & 8,9 \\
    Coma Bernices & 241.9 &  83.6 &   45 &   82.6 & 6,7 \\
            Draco &  86.4 &  34.7 &   92 & -104.0 & 8,10,11 \\
           Fornax & 237.3 & -65.6 &  140 &  -33.6 & 8,12,13 \\
         Hercules &  28.7 &  36.9 &  141 &  142.9 & 6,7 \\
              LMC & 280.5 & -32.9 &   49 &   73.8 & 8,14,15 \\
            Leo I & 226.0 &  49.1 &  257 &  179.0 & 8,16,17 \\
           Leo II & 220.2 &  67.2 &  235 &   26.5 & 8,18,19\\
           Leo IV & 265.4 &  56.5 &  154 &   13.9 & 6,7 \\
            Leo T & 214.9 &  43.7 &  422 &  -56.0 & 6,20 \\
            Leo V & 261.9 &  58.5 &  175 &   62.3 & 21 \\
              SMC & 302.8 & -44.3 &   60 &    9.0 & 8,22,23 \\
      Sagittarius &   5.6 & -14.1 &   16 &  166.3 & 8,24 \\
         Sculptor & 287.5 & -83.2 &   87 &   77.6 & 8,25,26 \\
          Segue 1 & 220.5 &  50.4 &   28 &  113.5 & 3,27 \\
          Segue 2 & 149.4 & -38.1 &   41 &   39.7 & 28 \\
          Sextans & 243.5 &  42.3 &   89 &   78.2 & 8,9,29 \\
     Ursa Major I & 159.4 &  54.4 &  101 &   -8.8 & 3,6 \\
    Ursa Major II & 152.5 &  37.4 &   36 &  -36.5 & 6,30 \\
       Ursa Minor & 104.9 &  44.8 &   77 &  -89.8 & 8,10,11 \\
        Willman 1 & 158.6 &  56.8 &   42 &   33.7 & 2,3 \\
\hline
\end{tabular}
\end{center}
\medskip
Sources: 1 - \citet{2006ApJ...647L.111B}, 2 - \citet{2007MNRAS.380..281M}, 3 - \citet{2008ApJ...684.1075M}, 4 - \citet{2009ApJ...690..453K}, 5 - \citet{2006ApJ...643L.103Z}, 6 - \citet{2007ApJ...670..313S}, 7 - \citet{2007ApJ...654..897B}, 8 - \citet{2004AJ....127.2031K}, 9 - \citet{1998ARA&A..36..435M}, 10 - \citet{2002AJ....124.3222B}, 11 - \citet{1995AJ....110.2131A}, 12 - \citet{2000A&A...355...56S}, 13 - \citet{2006AJ....131.2114W}, 14 - \citet{2001ApJ...553...47F}, 15 - \citet{2002AJ....124.2639V}, 16 - \citet{2004MNRAS.354..708B}, 17 - \citet{2007ApJ...657..241K}, 18 - \citet{2005MNRAS.360..185B}, 19 - \citet{2007AJ....134..566K}, 20 - \citet{2007ApJ...656L..13I}, 21 - \citet{2008ApJ...686L..83B}, 22 - \citet{2000A&A...359..601C}, 23 - \citet{2006AJ....131.2514H}, 24 - \citet{1997AJ....113..634I}, 25 - \citet{1995A&AS..112..407K}, 26 - \citet{1995A&A...300...31Q}, 27 - \citet{2009ApJ...692.1464G}, 28 - \citet{2009MNRAS.tmp..903B}, 29 - \citet{2006ApJ...642L..41W}, 30 - \citet{2006ApJ...650L..41Z}

\end{table}


\begin{table}
\begin{center}
  \caption{Data table for the satellites of M31.  Listed are Galactic
    coordinates ($l$, $b$) in degrees, actual distance $r$ from the
    centre of M31 in kpc, projected distance $R$ from the centre of
    M31 in kpc and corrected line-of-sight velocity in \kms. }
	\label{table:andsats}
	\begin{tabular}{lrrrrrc}
\hline
\hline
Name & \multicolumn{1}{c}{$l$} & \multicolumn{1}{c}{$b$} & \multicolumn{1}{c}{$r$} & \multicolumn{1}{c}{$R$} & \multicolumn{1}{c}{$\vlos$} & \multicolumn{1}{c}{Source} \\
 & \multicolumn{1}{c}{(deg)} & \multicolumn{1}{c}{(deg)} & \multicolumn{1}{c}{(kpc)} & \multicolumn{1}{c}{(kpc)} & \multicolumn{1}{c}{(\kms)} & \\
\hline
      M33 & 133.6 & -31.3 &  809 &  206 &    74 & 1,2 \\
      M32 & 121.1 & -22.0 &  785 &    5 &    95 & 2,3 \\
    IC 10 & 119.0 &  -3.3 &  660 &  261 &   -29 & 2,3,4 \\
  NGC 205 & 120.7 & -21.1 &  824 &   39 &    58 & 1,2 \\
  NGC 185 & 120.8 & -14.5 &  616 &  189 &   106 & 1,2 \\
  IC 1613 & 129.8 & -60.6 &  715 &  510 &   -56 & 2,3,5 \\
  NGC 147 & 119.8 & -14.2 &  675 &  144 &   117 & 1,2 \\
  Pegasus &  94.8 & -43.6 &  919 &  473 &    85 & 1,2 \\
   Pisces & 126.7 & -40.9 &  769 &  268 &   -37 & 1,2 \\
    And I & 121.7 & -24.8 &  745 &   59 &   -84 & 1,2 \\
   And II & 128.9 & -29.2 &  652 &  185 &    83 & 1,2 \\
  And III & 119.4 & -26.3 &  749 &   75 &   -57 & 1,2 \\
    And V & 126.2 & -15.1 &  774 &  109 &  -107 & 1,2 \\
   And VI & 106.0 & -36.3 &  775 &  267 &   -64 & 1,2 \\
  And VII & 109.5 &  -9.9 &  763 &  218 &    21 & 1,2 \\
   And IX & 123.2 & -19.7 &  765 &   41 &    94 & 1,6,7 \\
    And X & 125.8 & -18.0 &  702 &  110 &   130 & 8,9 \\
   And XI & 121.7 & -29.1 &  785 &  102 &  -140 & 7,10 \\
  And XII & 122.0 & -28.5 &  830 &  107 &  -268 & 7,10,11 \\
 And XIII & 123.0 & -29.9 &  785 &  115 &    64 & 7,10 \\
  And XIV & 123.0 & -33.2 &  740 &  161 &  -204 & 12 \\
   And XV & 127.9 & -24.5 &  770 &   94 &   -57 & 13,14 \\
  And XVI & 124.9 & -30.5 &  525 &  280 &  -106 & 13,14 \\
 And XVII & 120.2 & -18.5 &  794 &   45 &       & 15 \\
And XVIII & 113.9 & -16.9 & 1355 &  589 &       & 16 \\
  And XIX & 115.6 & -27.4 &  933 &  187 &       & 16 \\
   And XX & 112.9 & -26.9 &  802 &  128 &       & 16 \\
  And XXI & 111.9 & -19.2 &  859 &  148 &       & 17 \\
 And XXII & 132.6 & -34.1 &  794 &  220 &       & 17 \\
\hline
	\end{tabular}
\end{center}
\medskip
Sources: 
1 - \citet{2005MNRAS.356..979M}, 2 - \citet{2006MNRAS.365..902M}, 
3 - \citet{2004AJ....127.2031K}, 4 - \citet{1999ApJ...511..671S}, 
5 - \citet{1999AJ....118.1657C}, 6 - \citet{2004ApJ...612L.121Z}, 
7 - Collins et al. (2009, in prep), 8 - \citet{2007ApJ...659L..21Z}, 
9 - Kalirai et al. (2009, in prep), 10 - \citet{2006MNRAS.371.1983M}, 
11 - \citet{2007ApJ...662L..79C}, 12 - \citet{2007ApJ...670L...9M}, 
13 - \citet{2007ApJ...671.1591I}, 14 - \citet{2009arXiv0901.0820L}, 
15 - \citet{2008ApJ...676L..17I}, 16 - \citet{2008ApJ...688.1009M}, 
17 - \citet{2009arXiv0909.0399M}
\end{table}


\section{Mass Estimates for Andromeda and the Milky Way}
\label{sect:results}

\subsection{Choice of Power-Law Index Parameters}

We now apply the mass estimators to the Milky Way and M31, the two
largest galaxies in the Local Group. In converting heliocentric
quantities to Galactocentric ones, we assume a circular speed of
220~\kms at the Galactocentric radius of the sun ($R_{\odot}$ =
8.0~kpc) and a solar peculiar velocity of ($U,V,W$) =
(10.00,5.25,7.17)~\kms, where $U$ is directed inward to the Galactic
Centre, $V$ is positive in the direction of Galactic rotation at the
position of the sun, and $W$ is positive towards the North Galactic
Pole~\citep[see e.g.,][]{1998MNRAS.298..387D}.

For the M31 satellites, positional and velocity data must be computed
relative to M31 itself.  We take the position of M31 to be ($\ell$, b)
= (121.2$^\circ$, -21.6$^\circ$) at a distance of 785~kpc and its
line-of-sight velocity to be -123~\kms in the Galactic rest
frame (see e.g., McConnachie et al. 2005; McConnachie \& Irwin 2006).

\begin{figure}
\begin{center}
\includegraphics[width=0.5\textwidth]{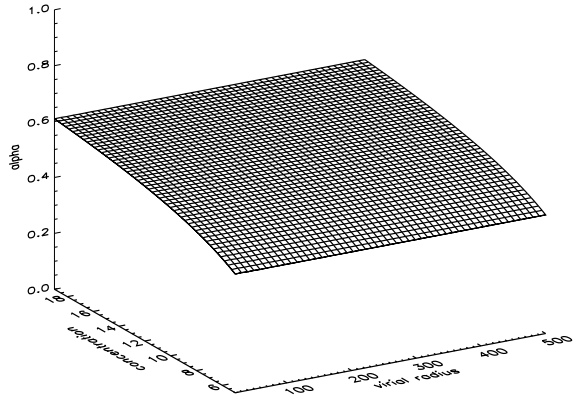}
\caption{The best-fit value of the power-law index $\alpha$ to an NFW
  profile as a function of the concentration and virial radius. Note
  that for plausible values of the concentration $c$ and the virial
  radius $\rvir$ for galaxies like the Milky Way and M31,
  $\alpha$ lies in the range 0.5-0.6.  The surface is smooth and
  flattish, implying that $\alpha$ is reasonably insensitive to the
  details of the NFW potential.}
	\label{fig:alpha_nfw}
\end{center}
\end{figure}

In order to apply our estimators to these systems, we need to compute
the power-law index of the host potential $\alpha$, the velocity
anisotropy $\beta$ and the power-law index of the satellite density
distribution $\gamma$.  There are cosmological arguments suggesting
that the potentials of dark haloes are well-approximated by
Navarro-Frenk-White (NFW) profiles \citep{1996ApJ...462..563N}.
Figure ~\ref{fig:alpha_nfw} shows the best-fit power-law to the NFW
potential for a wide range of concentrations and virial radii. The
fitting is performed in the region $10 < r/\mbox{kpc} < 300$, which is where
the majority of the satellites lie.  Now, \citet{2002ApJ...573..597K}
argued that the concentrations of the Milky Way and M31 are $c \approx
12$, whilst the virial radii $\rvir$ are in the range 250-300~kpc.
In other words, for the range of concentrations and virial radii
appropriate to galaxies like the Milky Way and M31, we see --
fortunately -- that the surface in Figure~\ref{fig:alpha_nfw} is
slowly-changing and flattish with $\alpha \approx 0.55$.

\begin{figure}
\begin{center}
	\includegraphics[width=0.4\textwidth]{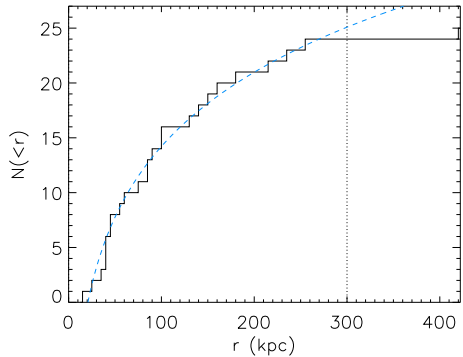}
	\includegraphics[width=0.4\textwidth]{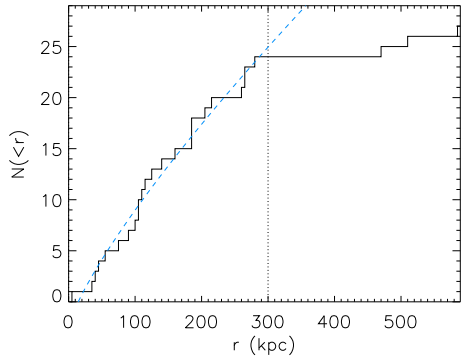}
	\caption{Cumulative numbers of satellite $N(<r)$ for the Milky Way
          (upper) and M31 (lower).  The best-fit power laws in the
          range $r \le 300\ \mbox{kpc}$ are also plotted.  The index of these
          power-law fits may be used to estimate the power-law index
          of the satellite density distribution $n(r)$.}
	\label{fig:nsats_r}
\end{center}
\end{figure}
\begin{figure}
\begin{center}
	\includegraphics[width=0.4\textwidth]{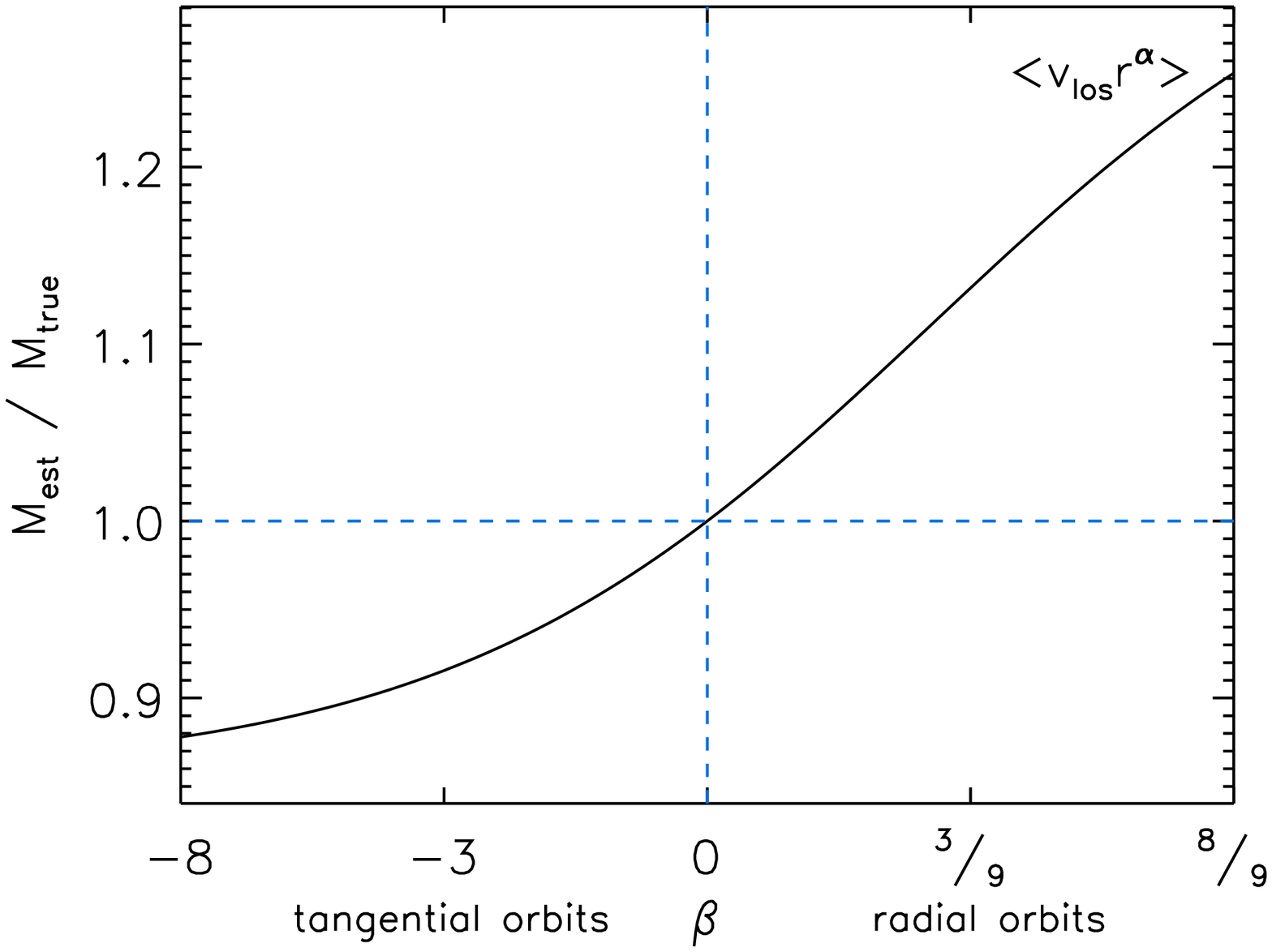}
	\includegraphics[width=0.4\textwidth]{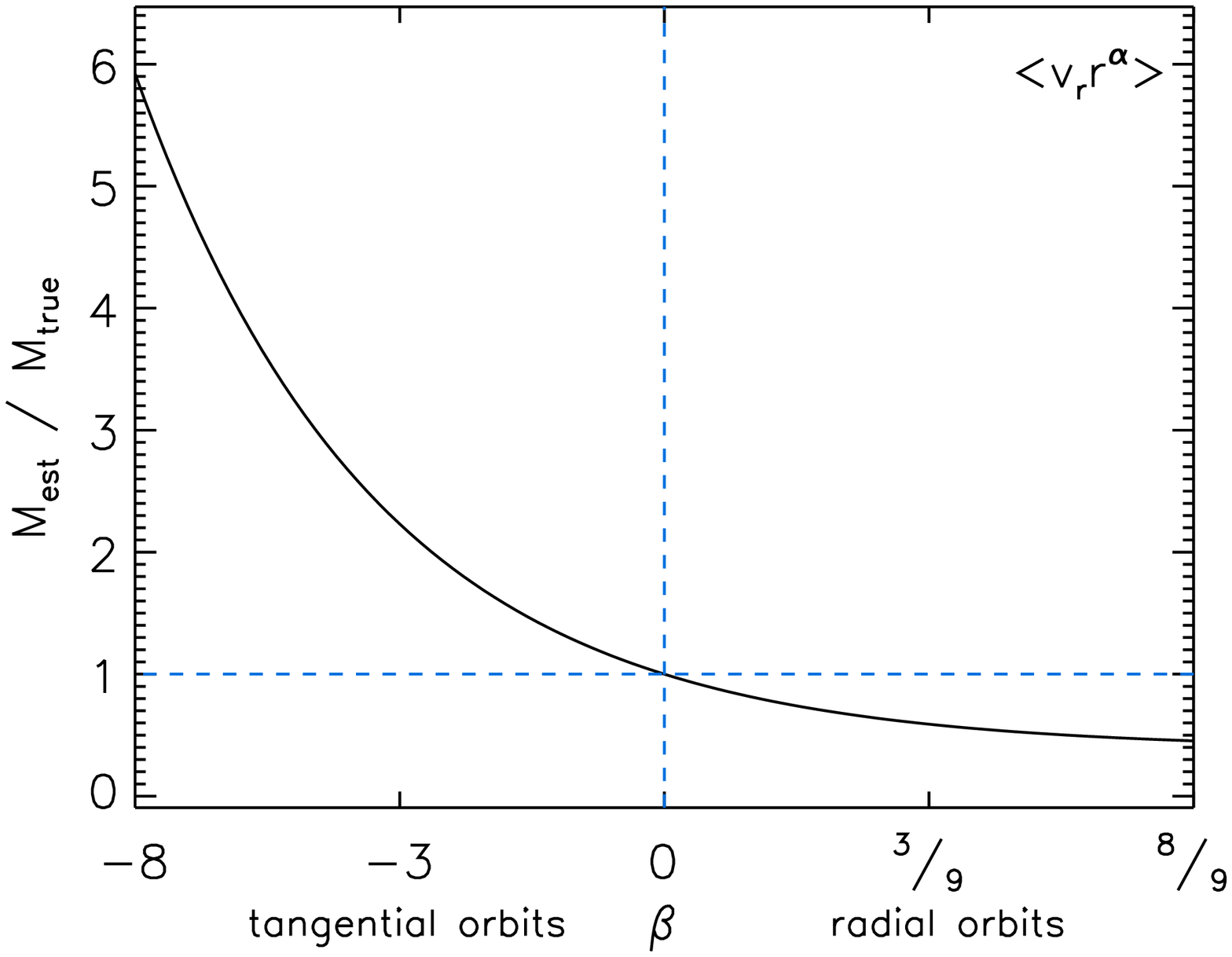}
        \caption{The sensitivity of the estimated mass on the
          anisotropy parameter $\beta$ for a satellite population with
          $\alpha= 0.55$, $\beta =0 $ and $\gamma = 2.7$. The figure
          shows the mass recovered using the input values of $\alpha$
          and $\gamma$ and varying the value of $\beta$. The
          functional form of the curve is easy to deduce. It is a
          rational function of $\beta$ for the upper panel, which uses
          the estimator of eqn~(\ref{eq:Andcase}), and a linear
          function of $\beta$ for the lower panel, which uses
          eqn~(\ref{eq:firstcase}).}
\label{fig:betauncertainty}
\end{center}
\end{figure}

If the satellite number density distribution $n(r)$ follows a power
law with index $\gamma$, then the number of satellites within any
radius, $N(< r)$, also follows a power-law with index $3-\gamma$.  We
fit power-laws to the Milky Way and M31 satellite cumulative
distributions in order to estimate $\gamma$. We restrict ourselves to
the inner regions of the satellite distributions, $r \le 300\
\mbox{kpc}$; beyond this range, the satellite population is likely to
be seriously incomplete. The distributions and the best-fitting power
laws are shown in Figure~\ref{fig:nsats_r}; the Milky Way data is
shown in the upper panel and M31 data is shown in the lower panel.  We
find $\gamma = 2.6$ for the Milky Way and $\gamma = 2.1$ for M31. Note
that data from the Sloan Digital Sky Survey
\citep[SDSS;][]{2000AJ....120.1579Y} has been instrumental in the
identification of many of the recently-discovered Milky Way dwarfs.
The SDSS coverage includes only the region around the North Galactic
Cap, and, as such, the distribution of known Milky Way satellites is
concentrated in that area of the sky. However, given our underlying
assumption that the distribution of satellites is spherically
symmetric, this directional bias does not impair our mass
estimators. A bigger worry may be the incompleteness in the satellite
distribution, which could affect the power index for the tracer number
density if the directional incompleteness varies in different
distances.

Finally, there are a number of possibilities for the velocity
anisotropy for the satellite galaxies. Previous studies often assumed
isotropy, arguing that there is no compelling evidence to the
contrary. However, \citet{2007ApJ...667..859D} found that the velocity
anisotropy of satellites in simulations behaves like $\beta(r) \simeq
0.55 \left( r /\rvir \right) ^{1/3}$ for $0.2 \rvir \le r \le \rvir$.
To estimate $\beta$ for the Milky Way and M31 satellites, we calculate
the weighted mean of this distribution
\begin{equation}
  \bar{\beta} = \frac{\int_{0.2\rvir}^{\rvir} \beta(r) n(r) r^2 dr}{\int_{0.2\rvir}^{\rvir} n(r) r^2 dr}
\end{equation}
where the weighting function $n(r)$ is the satellite number density
distribution. This gives $\bar{\beta} = 0.44$ for the Milky Way and
$\bar{\beta} = 0.45$ for M31. This is similar to the anisotropy of
halo stars ($\beta = 0.37$) in simulations reported by
~\citet{2008ApJ...684.1143X}.  Even though these numbers have the
backing of simulations, they are somewhat surprising.  Most of the
Milky Way satellites with measured proper motions are moving on polar
or tangential orbits. Using the sample of the 7 Milky Way satellites
with proper motions, we can compute the radial and tangential
components of the Galactocentric velocity.  From these, the observed
anisotropy $\beta \sim$ -4.5, which favours tangential orbits.  This
is consistent with the earlier, though indirect, estimate of
\citet{1999MNRAS.310..645W}, who found $\beta \sim$ -1, again
favouring tangential orbits. The origin of this discrepancy between
simulations and data is not well understood. Perhaps there is
considerable cosmic scatter in the anisotropy of the satellites, as it
may depend on the details of the accretion history of the host
galaxy. Figure~\ref{fig:betauncertainty} plots brings both good news
and bad. The upper panel shows that the mass estimates for external
galaxies using the line of sight estimator of eqn~(\ref{eq:Andcase})
are reasonably insensitive to the precise value of $\beta$. This make
sense, as for a galaxy like M31, the line of sight velocity encodes
information on both the radial and tangential velocity components
referred to the M31's centre. However, in the case of the Milky Way,
the situation is very different. The measured velocities provide
information almost wholly on the radial component referred to the
Galactic Center. In the absences of proper motions, the velocity
anisotropy is largely unconstrained by the data. This is the classical
mass-anisotropy degeneracy, and so -- as the lower panel shows --
there is considerable uncertainty in the mass estimates inferred using
eqn~(\ref{eq:firstcase}).

In what follows, we typically quote mass estimates for the
anisotropies derived both from observations $\beta_{\rm data}$ and
from simulations $\beta_{\rm sim}$, as well as for the case of
isotropy ($\beta =0$). In the absence of consistent indications to the
contrary, our preference is to assume isotropy and to give greatest
credence to the mass estimates obtained with this assumption.

\begin{table*}
\begin{center}
  \caption{Enclosed mass within 100, 200 and 300~kpc for the Milky
    Way and Andromeda galaxies.  We offer three estimates: one using
    the anisotropy inferred from data ($\beta \sim$ -4.5), one
    assuming isotropy ($\beta$ = 0) and the third with the anisotropy
    derived from simulations ($\beta \sim 0.45$).}
        \label{table:massests}
        \begin{tabular}{lccccccccc}
\hline
\hline
Galaxy & \multicolumn{3}{c}{$M_{300}$ ($\times 10^{11} \Msun$)} &
\multicolumn{3}{c}{$M_{200}$ ($\times 10^{11} \Msun$)} &
\multicolumn{3}{c}{$M_{100}$ ($\times 10^{11} \Msun$)} \\
 & $\beta_{\rm data}$ & isotropic & $\beta_{\rm sim}$ &
$\beta_{\rm data}$ & isotropic & $\beta_{\rm sim}$ &
$\beta_{\rm data}$ &
isotropic & $\beta_{\rm sim}$ \\
\hline
Milky Way               & 34.2 $\pm$ 9.3 &  9.2 $\pm$ 2.5 &  6.6 $\pm$ 1.8 & 21.
1 $\pm$ 5.7 &  5.5 $\pm$ 1.6 &  3.8 $\pm$ 1.0 & 13.9 $\pm$ 4.9 &  3.3 $\pm$ 1.1 
&  2.1 $\pm$ 0.7 \\
... excl Leo I          & 25.2 $\pm$ 7.5 &  6.9 $\pm$ 1.8 &  5.0 $\pm$ 1.2 &    
   ...      &       ...      &       ...      &       ...      &       ...      
&       ...      \\
... excl Leo I, Her     & 21.1 $\pm$ 6.3 &  5.8 $\pm$ 1.5 &  4.2 $\pm$ 1.1 & 17.
5 $\pm$ 5.3 &  4.6 $\pm$ 1.4 &  3.2 $\pm$ 0.8 &       ...      &       ...      
&       ...      \\
MW with PMs             & 38.6 $\pm$ 7.0 & 24.1 $\pm$ 5.3 & 22.1 $\pm$ 5.3 & 28.
3 $\pm$ 5.5 & 18.5 $\pm$ 4.2 & 17.0 $\pm$ 4.2 & 22.2 $\pm$ 5.0 & 13.8 $\pm$ 3.9 
& 11.4 $\pm$ 3.1 \\
... excl Draco          & 27.1 $\pm$ 4.9 & 14.1 $\pm$ 3.1 & 12.2 $\pm$ 2.7 & 18.
1 $\pm$ 3.4 & 10.0 $\pm$ 2.3 &  8.7 $\pm$ 2.2 & 12.9 $\pm$ 3.1 &  6.9 $\pm$ 1.9 
&  5.4 $\pm$ 1.5 \\
... excl LMC/SMC        & 38.8 $\pm$ 6.8 & 24.6 $\pm$ 5.8 & 21.7 $\pm$ 4.9 & 28.
3 $\pm$ 5.7 & 18.4 $\pm$ 4.3 & 16.3 $\pm$ 4.1 & 22.7 $\pm$ 5.6 & 13.9 $\pm$ 4.2 
& 11.2 $\pm$ 3.4 \\
... excl Draco, LMC/SMC & 25.9 $\pm$ 5.1 & 12.4 $\pm$ 2.9 & 10.6 $\pm$ 2.5 & 17.
0 $\pm$ 3.5 &  8.5 $\pm$ 2.2 &  7.2 $\pm$ 1.8 & 11.3 $\pm$ 2.9 &  5.4 $\pm$ 1.7 
&  4.2 $\pm$ 1.3 \\
\hline
M31                     & 15.8 $\pm$ 3.3 & 14.1 $\pm$ 4.1 & 13.1 $\pm$ 3.8 & 15.
4 $\pm$ 4.1 & 12.4 $\pm$ 3.8 & 10.6 $\pm$ 3.5 &  2.6 $\pm$ 1.0 &  2.1 $\pm$ 1.0 
&  1.8 $\pm$ 1.0 \\
... excl AndXII         & 12.2 $\pm$ 2.7 & 10.9 $\pm$ 3.1 & 10.1 $\pm$ 3.2 & 11.
4 $\pm$ 3.2 &  9.2 $\pm$ 3.1 &  7.9 $\pm$ 2.8 &       ...      &       ...      
&       ...      \\
... excl AndXII, AndXIV &  9.6 $\pm$ 2.1 &  8.5 $\pm$ 2.4 &  8.0 $\pm$ 2.4 &  8.
6 $\pm$ 2.6 &  6.9 $\pm$ 2.4 &  5.9 $\pm$ 2.2 &       ...      &       ...      
&       ...      \\
M31 with PMs            & 15.1 $\pm$ 3.8 & 13.9 $\pm$ 3.5 & 13.1 $\pm$ 3.5 &    
   ...      &       ...      &       ...      &       ...      &       ...      
&       ...      \\
\hline
        \end{tabular}
\end{center}
\end{table*}

\begin{figure}
\begin{center}
	\includegraphics[width=0.47\textwidth]{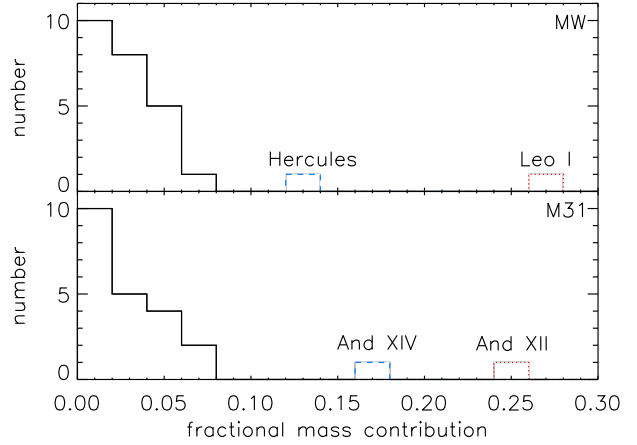}
	\caption{The fractional contribution each satellite makes to
          the mean mass estimator for the Milky Way (top) and M31
          (bottom).  For both galaxies, the mass budget is dominated
          by two satellites.  For the Milky Way these are Leo I (red,
          dotted) and Hercules (blue, dashed).  For M31, these are And
          XII (red, dotted) and And XIV (blue, dashed).}
	\label{fig:massconts}
\end{center}
\end{figure}

\subsection{Radial Velocity Datasets}

Armed with values for $\alpha$, $\beta$ and $\gamma$, we now set the
mass estimators to work.  Data for the satellites of the Milky Way and
M31 are given in Tables~\ref{table:mwsats} and ~\ref{table:andsats}
respectively. Objects for which no line-of-sight velocity has been
measured (And XVII, And XVIII, And XIX, And XX, And XXI and And XXII)
are included in the tables, but excluded from the analysis.

Using eqn~(\ref{eq:firstcase}) and recalling that the Monte Carlo
simulations gave errors of $\sim25\%$, we give estimates of the mass
with 100, 200 and 300~kpc for the Milky Way Galaxy in
Table~\ref{table:massests}. Assuming velocity isotropy, we obtain for
the mass of the Milky Way $M_{300} = 0.9 \pm 0.3 \times 10^{12}
\Msun$.  The cussedness of the mass-anisotropy degeneracy is well
illustrated by the fact that using the observationally derived
$\beta_{\rm data}$ gives $M_{300} = 3.4 \pm 0.9 \times 10^{12} \Msun$,
whilst using that from simulations gives $M_{300} = 0.6 \pm 0.2 \times
10^{12} \Msun$. The huge spread in mass estimates is due to the fact that
the line-of-sight velocities for the satellites are almost entirely
providing information on the radial velocities as judged from the
Galactic Centre.  There is almost no information on the tangential
motions in our dataset. However, there are other astrophysical reasons
why masses higher than $\sim 2 \times 10^{12} \Msun$ are disfavoured.

Using eqn~(\ref{eq:Andcase}), we obtain the mass of M31 within 300~kpc
as $M_{300} = 1.4 \pm 0.4 \times 10^{12} \Msun$. Here, though, in
sharp distinction to the case of the Milky Way, plausible changes in
the velocity anisotropy generate modest changes of the order of 10 per
cent in the mass estimate, as shown in Table~\ref{table:massests}. Of
course, this is understandable, as the line-of-sight velocity now has
information on both the radial and tangential components, albeit
tangled up in the projection.

Taking the masses derived using velocity isotropy ($\beta=0$), we note
that this work hints at the removal of a long-standing puzzle, namely
that the kinematic data on the satellite galaxies suggested that M31
was less massive than the Milky Way, whereas other indicators (such as
the total numbers of globular clusters or the amplitude of the gas
rotation curve) suggested the reverse. In fact, with the new datasets,
the ratio of the masses of M31 to the Milky Way ($\sim 1.5$) is close
to that which would be inferred using the Tully-Fisher relationship
and the assumption that the luminosity is proportional to the total
mass ($250^4/220^4 \approx 1.67$). If instead the radial anisotropies
derived from simulations are preferred, then the ratio is $\sim 1.98$.

However, it may be imprudent to include all the satellites. For
example, Leo I has long been known to dominate mass estimates of the
Milky Way, on account of its large distance ($\sim 260\ \mbox{kpc}$) and high
line-of-sight velocity~\citep[see
e.g.][]{1992MNRAS.255..105K,1996ApJ...457..228K,1999MNRAS.310..645W}.
It is unclear that Leo I is actually on a bound orbit, as opposed to a
hyperbolic one.  Hence, many attempts at determining the mass of the
Milky Way quote estimates both including and excluding Leo I.  In
fact, recent photometric and spectroscopic evidence presented
by~\citet{2007ApJ...663..960S} favours the picture in which Leo I is
bound on an orbit with high eccentricity ($\sim 0.95$) and small
perigalacticon (10-15~kpc). In particular, such models give good
matches to the surface density and radial velocity dispersion profiles
of Leo~I, and imply high mass estimates for the Milky Way. However,
\citet{2007MNRAS.379.1475S} using simulations found a population of
satellite galaxies on extreme orbits ejected from haloes as a result
of three-body slingshot effects, and suggested that Leo I might be an
example of such an object.  So, although the present evidence favours
a bound orbit, a definitive verdict must await the measurement of
Leo~I's proper motion by the {\it Gaia} satellite, which should
resolve the issue.

\begin{figure*}
\begin{center}
	\includegraphics[width=0.7\textwidth]{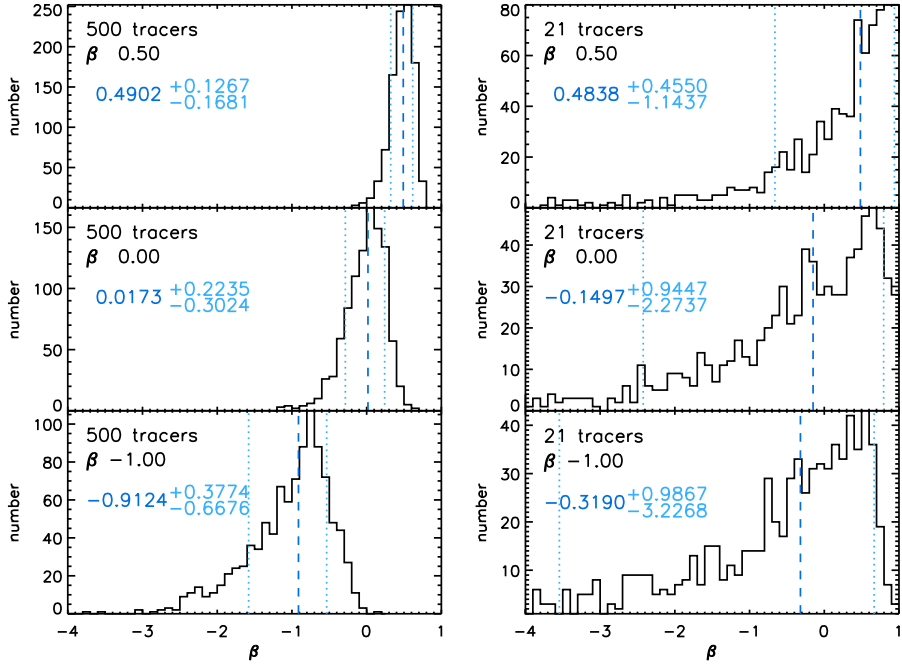}
	\caption{Distribution of $\beta$ obtained from simultaneous 
          mass and velocity anisotropy fitting. The median of each 
          distribution is shown as a blue dashed line and the 
          68 $\%$ confidence limits as cyan dotted lines. These values 
          are also given in blue and cyan on the plot. The left 
          panel shows the idealised case for 500 tracers, the right 
          panel the case tailored to our M31 data where only 21
          tracers are used.  The value of $\beta$ used to generate 
          the tracer population is also shown.}
	\label{fig:simulterrors}
\end{center}
\end{figure*}

Given that there is one satellite that is known to inflate the Milky
Way's mass, it is interesting to investigate whether any of the other
satellites, particularly the recent discoveries, play similar
r\^{o}les.  The upper panel of Figure \ref{fig:massconts} shows the
fractional contributions each satellite makes to the Milky Way's mass
($C \vlos^2r^{\alpha} / (G N) $) -- it is the total of these values
that we take to be the mass estimate.  There are two clear outliers;
the outermost satellite in this distribution is Leo I, the less
extreme satellite is Hercules.  Like Leo I, Hercules has a substantial
radial velocity and a relatively large Galactocentric distance ($\sim
130\ \mbox{kpc}$). Hercules has a highly elongated, irregular and
flattened structure~\citep{2007ApJ...654..897B,
  2007ApJ...668L..43C}. This is consistent with tidal disruption
during pericentric passages on a highly eccentric orbit ($e >
0.9$). This seems good evidence that Hercules is truly bound to the
Milky Way.

We repeat the same analysis for M31 and the results are shown in the
bottom panel of Figure \ref{fig:massconts}.  Interestingly, we see
that there are two outliers in the distribution, namely two of the
recent discoveries, And XII and And XIV. Notice that though both
objects have a substantial effect on M31's mass estimate, neither are
as extreme as Leo I.  It is the inclusion of these two new objects in
the satellite dataset that has augmented the mass of M31, so that it is
now somewhat greater than that of the Milky Way.

But, this begs the question: should these satellites be included? And
XIV was discovered by \citet{2007ApJ...670L...9M} in a survey of the
outer M31 stellar halo.  They recognized its extreme dynamical
properties and suggested that it may either be falling into M31 for
the first time or that M31's mass must be larger than hitherto
estimated by virial arguments. In fact, And XIV's lack of gas and its
elongated structure suggest that ram pressure stripping and tidal
effects may have been important in its evolution. This is consistent
with And XIV being a true satellite of M31 that has already suffered a
pericentric passage, a conclusion that could be strengthened with
deeper imaging, which might reveal the presence of tidal tails around
And XIV.

And XII is a still more ambiguous object -- it was discovered as a
stellar overdensity by \citet{2006MNRAS.371.1983M}. Spectroscopic
observations were subsequently taken by \citet{2007ApJ...662L..79C},
who conjectured that the satellite might be falling into the Local
Group for the first time. The evidence for this is its large velocity
and its likely location behind M31. However, it remains unclear
whether this evolutionary track is consistent with the absence of
detection of HI gas in the object. Pristine, infalling dwarfs, which
have not yet experienced a pericentric passage of 50~kpc or less,
should retain sizeable amounts of neutral HI gas, whereas
\citet{2007ApJ...662L..79C} constrain the mass in HI to be less than
$3 \times 10^3 \Msun$.

In light of this, we provide more mass estimates, after removing
possible ambiguous objects (and re-computing the parameter $\gamma$
where necessary). For the Milky Way, we exclude Leo I only and then
Leo I and Hercules. For M31, we exclude And XII only and then And XII
and And XIV.  These mass estimates are also shown in Table
\ref{table:massests}. Note that, for example, the exclusion of Leo I
does not change the mass estimate within 100 or 200~kpc, as Leo I is
outside of this range. Similarly, And XII and And XIV lie outside of
100~kpc from the center of M31, so the mass estimates without them do
not change the final column of the table.

In the case of velocity isotropy ($\beta=0$), it requires the excision
of both And XII and And XIV from the datasets for the mass estimate of
M31 to become comparable to or smaller than the Milky Way. For
example, the mass of M31 with And XII and And XIV both removed is
$0.85 \pm 0.24 \times 10^{12} \Msun$, as compared to the mass of the
Milky Way with Leo I retained of $0.92 \pm 0.25 \times 10^{12}
\Msun$. However, we have argued that And XIV is most likely bound,
whilst And XII is a more ambiguous case. In other words, the problem
pointed to by \citet{2000MNRAS.316..929E} -- namely that the mass of
M31 inferred from the kinematics of the satellites is less than the
mass of the Milky Way -- has indeed been ameliorated by the discovery of
more fast-moving M31 satellites.

It seems particularly intriguing that such satellites exist for both
the Milky Way and M31. \citet{1999MNRAS.310..645W} used virialized
models to estimate that the probability that, in a sample of 30
satellites, there is an object like Leo~I, which changes the mass
estimate by a factor of a few. They found that the probability is
minute, only $\sim 0.5 \%$. Prior expectation does not favour the
existence of objects like Leo I or And XII, yet in fact, both big
galaxies in the Local Group possess such satellites. The clear
conclusion is that the satellites in the outer parts of these galaxies
cannot all be virialized. This is a point in favour of processes such
as those advocated by \citet{2007MNRAS.379.1475S} to populate such
orbits.

\begin{table}
\begin{center}
  \caption{Table of proper motion data for the satellites of the Milky
    Way and M31.  Listed are equatorial proper motions in mas
    century$^{-1}$.}
        \label{table:pms}
\begin{tabular}{lccc}
\hline
\hline
Name & \multicolumn{1}{c}{$\mu_{\alpha} \cos \delta$} &
\multicolumn{1}{c}{$\mu_{\delta}$} & \multicolumn{1}{c}{Source} \\
 & \multicolumn{1}{c}{(mas/century)} & \multicolumn{1}{c}{(mas/century)} & \\
\hline
           Carina &   22 $\pm$   9 &   15 $\pm$   9 &  1 \\
            Draco &   60 $\pm$  40 &  110 $\pm$  30 &  2 \\
           Fornax &   48 $\pm$   5 &  -36 $\pm$   4 &  3 \\
          LMC/SMC &  198 $\pm$   5 &   25 $\pm$   5 &  4 \\
         Sculptor &    9 $\pm$  13 &    2 $\pm$  13 &  5 \\
          Sextans &  -26 $\pm$  41 &   10 $\pm$  44 &  6 \\
       Ursa Minor &  -50 $\pm$  17 &   22 $\pm$  16 &  7 \\
              M33 &  2.1 $\pm$ 0.7 &  2.5 $\pm$ 1.2 &  8 \\
             IC10 & -0.2 $\pm$ 0.8 &  2.0 $\pm$ 0.8 &  9 \\
\hline
              M31 &  2.1 $\pm$ 1.1 & -1.0 $\pm$ 0.9 &  10 \\
\hline
\end{tabular}
\end{center}
\medskip
Sources: 1 - \citet{2003AJ....126.2346P}, 2 - \citet{1994IAUS..161..535S}, 3 -
\citet{2007AJ....133..818P}, 4 - \citet{2008AJ....135.1024P}, 5 -
\citet{2006AJ....131.1445P}, 6 - \citet{2008ApJ...688L..75W}, 7 -
\citet{2005AJ....130...95P}, 8 - \citet{2007arXiv0708.1704B}, 
9 - \citet{2007A&A...462..101B}, 10 - \citet{2008ApJ...678..187V},
though unlike the other proper motions, this not a measurement but 
inferred from indirect evidence.
\end{table}
\begin{figure}
\begin{center}
	\includegraphics[width=0.4\textwidth]{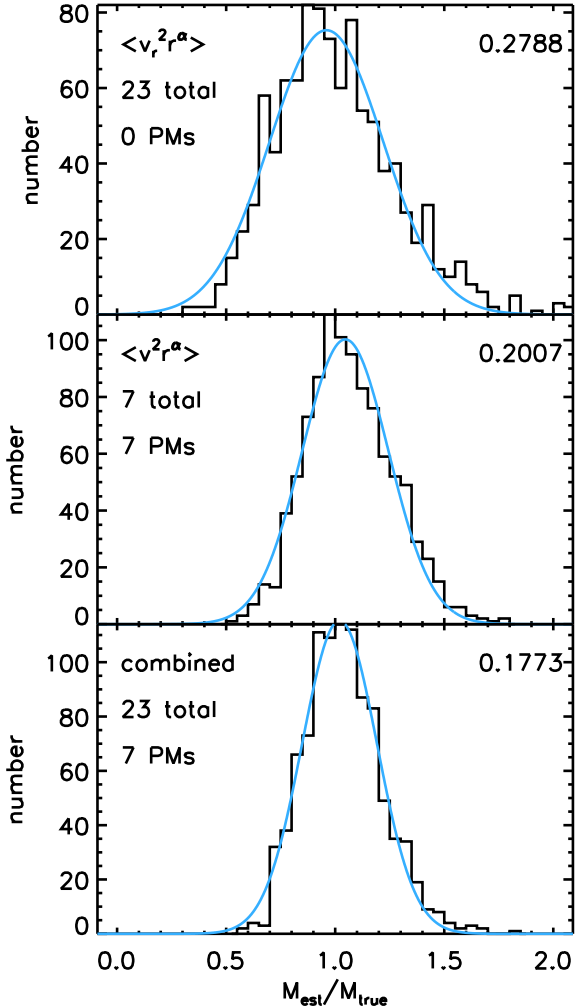}
	\caption{Distribution of mass estimates as a fraction of true
          mass for Monte Carlo simulations using (top) 23 satellites
          with radial velocities, (middle) 7 satellites with proper
          motions and (bottom) 23 satellites, 7 of which have proper
          motions. The standard deviation of the best fitting Gaussian
          is shown in the top-right hand corner of each panel. [These
          plots assume $\beta = -4.51$, as estimated from the data].}
	\label{fig:milkywaycase}
\end{center}
\end{figure}
\begin{figure}
\begin{center}
	\includegraphics[width=0.48\textwidth]{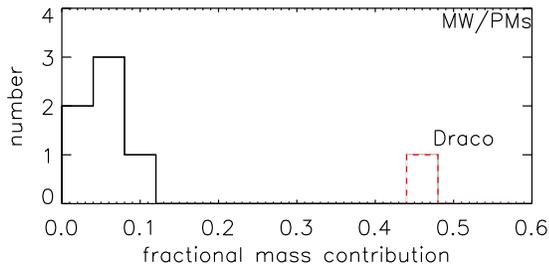}
	\caption{The fractional contribution each satellite with
          proper motions makes to the mean mass estimate for the Milky
          Way Galaxy. Notice the extreme effect of Draco's proper
          motion.}
	\label{fig:pmconts}
\end{center}
\end{figure}


\subsection{Simultaneous Solution for Mass and Anisotropy}

There is one further way in which the estimators can be set to work
with the line-of-sight velocities. When three dimensional positions
and projected positions are simultaneously available -- as for example
in the case of M31's satellites -- it is possible to use the
estimators based on both the $\langle \vlos^2 r^\alpha \rangle$ and
the $\langle \vlos^2 R^\alpha \rangle$ moments to solve simultaneously
for both the total mass and the anisotropy parameter. There is however
no guarantee that the solution for $\beta$ is in the physical range
$-\infty \le \beta \le 1$.

The success of this procedure of course rests on the accuracy of the
data. The distances of the M31 satellites are determined by the tip of
the red giant branch method and have errors of $\pm 30$ kpc \citep[see
  e.g.,][]{2005MNRAS.356..979M}. If we use eqns~(\ref{eq:Andcase}) and
(\ref{eq:lastcase}), and simultaneously solve for the unknowns, we
obtain 
\begin{equation}
M_{300} = 1.5 \pm 0.4 \times 10^{12} \Msun, \qquad\qquad \beta = -0.55^{+1.1}_{-3.2}
\end{equation}
which corresponds to mild tangential anisotropy. These are
surprisingly sensible answers given the distance errors.

Fig.~\ref{fig:simulterrors} is inferred from Monte Carlo simulations
and shows the distributions of anisotropy parameters derived from
simultaneous mass and anisotropy fitting for mock datasets. Also given
in the panels are the median and 68 per cent confidence limits for the
anisotropy parameter, in the case of 21 satellite galaxies
(comparable to the present dataset for M31) and the case of 500
satellites. Although with 21 tracers, the errors on the anisotropy
parameter are substantial, matters improve significantly with larger
numbers of tracers. A dataset of 500 halo satellites (dwarf galaxies,
globular clusters and planetary nebulae) is not unreasonable for a
galaxy like M31 in the near future.  This raises the possibility that
the method of simultaneous fitting may prove more compelling in the
future. In fact, given 500 tracers, it is reasonable to use the
estimators based on both the $\langle \vlos^2 r^\alpha \rangle$ and
the $\langle \vlos^2 R^\alpha \rangle$ moments to fit simultaneously
at each distance, thus giving the run of anisotropy parameter and mass
with radius.

\subsection{Radial and Proper Motion Datasets}

\label{sect:addingpms}
Thus far, we have used only the line of sight velocities to make mass
estimates. In this section, we add in the proper motions of
satellites, where available.  Thus, for the Milky Way galaxy, we
combine results from eqn~(\ref{eq:firstcase}) for satellites without
proper motions and from eqn~(\ref{eq:PMcase}) for those with proper
motions, weighting each estimate by the reciprocal of the standard
deviation to give the final answer.

Proper motions, albeit with large error bars, have been measured for a
total of 9 of the Milky Way satellite galaxies.  It seems prudent to
exclude Sagittarius, as it is in the process of merging with the Milky
Way. Additionally, the interacting Magellanic Clouds are treated as a
single system by computing the proper motion of their mass centroid,
taking the masses of the LMC and SMC as $\sim 2 \times 10^{10} \Msun$
and $2 \times 10^9 \Msun$ respectively~\citep{1997NewA....2...77K}.
This leaves us with a set of 7 satellites with proper motion data,
summarized in Table \ref{table:pms}.  In most cases, errors on proper
motions are large and, where multiple studies exist, the measurements
are sometimes in disagreement.  The proper motions inferred by ground
based methods are in reasonable agreement with those derived from the
{\it Hubble Space Telescope} ({\it HST}) in the cases of Fornax
~\citep{2007AJ....133..818P,2008ApJ...688L..75W},
Carina~\citep{2003AJ....126.2346P,2008ApJ...688L..75W} and the
Magellanic
Clouds~\citep{2008AJ....135.1024P,2009AJ....137.4339C}. But, for Ursa
Minor~\citep{1994IAUS..161..535S,2005AJ....130...95P} and for Sculptor
\citep{2006AJ....131.1445P,2008ApJ...688L..75W}, agreement between
different investigators is not good, and we have preferred to use the
estimates derived from {\it HST} data.  Nonetheless, it is important
to include the proper motion data, especially for mass estimates of
the Milky Way Galaxy.  We use these proper motions along with distance
and line-of-sight velocity data to calculate full space velocities for
these satellites, as described in \citet{2002AJ....124.3198P}.

In addition, there are two satellites of M31 with measured proper
motions, namely M33 and IC 10. This astonishing feat has exploited the
{\it Very Long Baseline Array} to measure the positions of water
masers relative to background quasars at multiple epochs
\citep{2005Sci...307.1440B,2007A&A...462..101B}. Unfortunately, the
technique cannot be extended to M31 itself, as it does not contain any
water masers, and so its proper motion is much less securely
known. However, \citet{2008ApJ...678..187V} reviewed the evidence from
a number of sources -- including kinematics of the M31 satellites, the
motions of the satellites at the edge of the Local Group, and the
constraints imposed by the tidal distortion of M33's disk -- to
provide a best estimate. These data are also listed in
Table~\ref{table:pms}.

The Milky Way satellites are so remote that their line-of-sight
velocities in the Galactic rest frame are almost identical to their
radial velocities, as judged form the Galactic Centre. The proper
motion data provide genuinely new information on the tangential
motions and this is the only way to break the mass-anisotropy
degeneracy. The same argument does not hold with equal force for M31,
as the line of sight velocities incorporate contributions from both
the radial and tangential components as reckoned from the centre of
M31. Nonetheless, it is good practice to use all the data available,
even though the proper motions of M33 and IC 10 with respect to the
M31 reference frame must be inferred using an estimate of M31's proper
motion (rather than a measurement).

For the satellites without proper motions, we use the form of the
estimator given in eqns~(\ref{eq:firstcase}) or (\ref{eq:Andcase}) for
the Milky Way and M31 respectively; for those with proper motions, we
use eqn~(\ref{eq:PMcase}). We combine results from the two estimators,
weighting each estimate by the reciprocal of the standard deviation to
give the final answer. To infer the standard deviation, we perform
Monte Carlo simulations. So, for the case of the Milky Way, we
generate mock datasets of 25 satellites, for which only 7 have proper
motions.  The errors on radial velocities are dwarfed by the
uncertainty caused by the small number statistics and so are
neglected. But, the errors on the proper motions are not negligible
and they are incorporated into the simulations by adding a value
selected at random from the range [-0.5 $\mu$, 0.5 $\mu$], where $\mu$
is the proper motion. The flat distribution has been chosen as
systematic errors are as important as random Gaussian error in the
determination of proper motions. However, we have tested alternatives
in which we use the relative observational errors, or the relative
observational errors multiplied by 2.5, and find that our results are
robust against changes to the error law.  The standard deviations of
the fractional mass distribution the satellites with and without
proper motions are separately computed, as illustrated in the panels
of Figure~\ref{fig:milkywaycase}. We linearly combine the mass
estimates, weighting with the reciprocal of the standard deviation, to
give the final values reported in Table~\ref{table:massests}.

Given that the Milky Way satellites with measured proper motions are
moving on polar orbits, it is no surprise that the mass estimate of the
Milky Way has now increased.  Adopting the value of $\beta$ we estimate from 
the data, we find $M_{300} = 3.9 \pm 0.7 \times 10^{12} \Msun$ for the Milky 
Way Galaxy and $M_{300} = 1.5 \pm 0.4 \times 10^{12} \Msun$ for M31. Assuming  
isotropy, we find $M_{300} = 2.5 \pm 0.5 \times 10^{12} \Msun$ for the Milky 
Way Galaxy and $M_{300} = 1.4 \pm 0.4 \times 10^{12} \Msun$ for M31. Notice 
however, the mass estimate for M31 has barely changed from the value inferred
from the full radial velocity dataset.

Again, we calculate the contribution that each satellite makes to the
mass estimate to investigate whether any are dominating the final
answer.  First, this procedure guards against the possibility of a
completely rogue proper motion measurement.  Second, there are some
suggestions that the Magellanic Clouds may not be bound, or even if
bound may only be on its second passage and so may not be part of the
relaxed distribution of satellite galaxies
~\citep{2007ApJ...668..949B}. So, it is helpful to check that our
results are not unduly sensitive to its inclusion.  As
Figure~\ref{fig:pmconts} shows, we find that Draco is a clear outlier
and nearly doubles the Milky Way mass estimate.  If we remove the
Draco proper motion from the sample, we instead recover a mass
$M_{300}$ = 2.7 $\pm$ 0.5 $\times 10^{12} \Msun$ (assuming
$\beta_{\rm data}$) or $M_{300}$ = 1.4 $\pm$ 0.3 $\times 10^{12} \Msun$
(assuming isotropy). It is particularly concerning that the proper
motion of Draco has such a substantial effect, because -- as judged
from the size of the error bars in Table~\ref{table:pms} -- it is one
of the noisier measurements. By contrast, the exclusion of the
Magellanic Clouds has only a minor effect, as is evident from the
results listed in Table~\ref{table:massests}.

We have covered a number of possibilities, so it is probably useful
for us to give our best estimates. On balance, we think the case for
including at least And XIV among the satellite sample for Andromeda is
strong. Whilst And XII is a more ambiguous case, the lack of any HI
gas suggests to us that it should also be included. Among the
satellites of the Milky Way, we favour including Leo I based on the
work of~\citet{2007ApJ...663..960S}, whilst we are inclined to discard
the proper motion of Draco reported in~\citet{1994IAUS..161..535S}
until corroborated. Until the discrepancy between the velocity
anisotropies reported in simulations and in data is explained, we
prefer to use the data as our guide 

{\it So, our best estimate for the mass of the Milky Way within
  300~kpc is
\begin{equation}
M_{300} \sim 2.7 \pm 0.5 \times 10^{12} \Msun
\label{eq:MWMASS}
\end{equation} 
whilst for M31, it is 
\begin{equation}
M_{300}\sim
1.5 \pm 0.4 \times 10^{12} \Msun.
\label{eq:ANDMASS}
\end{equation}
} 
These estimates are obtained using the combined radial velocity and
proper motion datasets. The error bars only incorporate the
statistical uncertainty. As we have emphasised, there are much greater
uncertainties induced by selection of satellite members and velocity
anisotropy. In particular, when these uncertainties are considered, it
is not possible to decide which of the Milky Way or M31 is more
massive based on satellite kinematic data alone.

\section{Discussion}

It is instructive to compare our results with a number of recent
estimates of the masses of the Local Group and its component galaxies.
\citet{2008ApJ...684.1143X} extracted a sample of $\sim 2400$ blue
horizontal branch stars from the SDSS. These are all resident in the
inner halo within 60~kpc of the Galactic centre. This has the
advantage that the BHBs are surely virialized, but the disadvantage
that no inference can be made about the mass exterior to 60~kpc.
Hence, any estimate as to the total mass is driven wholly by
prior assumptions rather than the data. In fact,
\citet{2008ApJ...684.1143X} assumed an NFW halo with a canonical
concentration holds, and then estimated the virial mass of the Milky
Way's dark matter halo as $M = 1.0^{+0.3}_{-0.2} \times 10^{12} \Msun$,
using Jeans modelling with an anisotropy parameter inferred from
numerical simulations. This is lower than our preferred value, but in
good agreement with our comparable calculations using line of sight
velocity datasets alone.

A somewhat similar calculation for M31 has been reported by
\citet{2008MNRAS.389.1911S}. The mass of the baryonic material is
estimated using a Spitzer 3.6 $\mu$m image of the galaxy, together
with a mass-to-light ratio gradient based on the galaxy's $B - R$
colour. This is combined with an adiabatically-contracted NFW halo
profile to reproduce the observed HI rotation curve data. They find a
total virial mass of M31's dark halo as $8.2 \pm 0.2 \times 10^{11}
\Msun$. This is lower than most of our estimates, with the exception
of those based on samples excluding both And XII and And XIV.

Although these calculations are interesting, it is worth remarking
that the final masses are not wholly controlled by the data. We know
that, from Newton's theorem, any mass distribution outside the
limiting radius of our data has no observational effect in a spherical
or elliptical system. To estimate the virial mass from data confined
to the inner parts (such as BHBs or the optical disk) requires an
understanding of the structure of the pristine dark halo initially, as
well as how it responds to the formation of the luminous baryonic
components. It is this that controls the final answer.

\citet{2008MNRAS.384.1459L} used the Millennium Simulation to extract
mock analogues of the Local Group and calibrate the bias and error
distribution of the Timing Argument estimators~\citep[see
e.g.,][]{1959ApJ...130..705K,1989MNRAS.240..195R}. From this, they
obtain a total mass of the two large galaxies in the Local Group of
$5.3 \times 10^{12} \Msun$ with an inter-quartile range of [$3.8 \times
10^{12}, 6.8 \times 10^{12}$] $\Msun$ and a 95 \% confidence lower
limit of $1.8 \times 10^{12} \Msun$. Importantly,
\citet{2008MNRAS.384.1459L} showed that the mass estimate from the
timing argument is both unbiased and reasonably robust. This is a
considerable advance, as there have long been worries that the gross
simplification of two-body dynamics implicit in the original
formulation of the Timing Argument may undermine its conclusions.

It therefore seems reasonable to assume that the combined mass of the
Milky Way Galaxy and M31 is at least $3.8 \times 10^{12} \Msun$, and
perhaps more like $5.3 \times 10^{12} \Msun$. The low estimates of the
Milky Way and M31 masses of \citet{2008ApJ...684.1143X} and
\citet{2008MNRAS.389.1911S} are not compatible with this, and barely
compatible with Li \& White's 95 \% lower limit.  Using our preferred
values in eqns~(\ref{eq:MWMASS}) and (\ref{eq:ANDMASS}), the combined
mass in the Milky Way and M31 galaxies is $4.2 \pm 0.6 \times 10^{12}
\Msun$. This is comparable to the $3.8 \times 10^{12} \Msun$ of Li \&
White.

\citet{2008MNRAS.384.1459L} also estimated a virial mass for the Milky
Way of $2.4 \times 10^{12} \Msun$ with a range of [$1.1 \times
10^{12}, 3.1 \times 10^{12}$] $\Msun$, based on timing arguments for
Leo I. Given all the uncertainties, this is in remarkable accord with
our best estimate.


\section{Conclusions}
\label{sect:conclusions}

We have derived a set of robust tracer mass estimators, and discussed
the conditions under which they converge. Given the positions and
velocities of a set of tracers -- such as globular clusters, dwarf
galaxies or stars -- the estimators compute the enclosed mass within
the outermost datapoints. The accuracy of the estimator has been
quantified with Monte Carlo simulations. The estimators are applicable
to a wide range of problems in contemporary astrophysics, including
measuring the masses of elliptical galaxies, the haloes of spiral
galaxies and galaxy clusters from tracer populations. They are
considerably simpler to use than distribution function based
methods~\citep[see
e.g.][]{1987ApJ...320..493L,1992MNRAS.255..105K,1999MNRAS.310..645W},
and involve no more calculation than taking weighted averages of
combinations of the positional and kinematical data. They should find
widespread applications.

The mass estimators are applied to the satellite populations of the
Milky Way and M31 to find the masses of both galaxies within 300~kpc.
These estimates are the first to make use of the recent burst of
satellite discoveries around both galaxies. Both satellite populations
have nearly doubled in size since previous estimates were made.  We
summarise our results by answering the questions; What are (1) the
minimum, (2) the maximum and (3) the most likely masses of the Milky
Way and M31 galaxies?

\medskip
\noindent
(1) The mass of the Milky Way Galaxy within 300~kpc could be as low as
$0.4\pm 0.1 \times 10^{12} \Msun$. This would imply that Leo I is
gravitationally unbound, contrary to the recent evidence provided by
by~\citet{2007ApJ...663..960S}.  Leo I would then be either an
interloper or an object being ejected from the Milky Way by an
encounter. It would also require that the proper motion of Draco
\citep{1994IAUS..161..535S} is incorrect, which is not inconceivable
given the difficulty of the measurements. It implies that the
satellite galaxies are moving on radial orbits and so the velocity
anisotropy is radial.

The mass of M31 within 300~kpc could plausibly be as low as $0.8 \pm
0.2 \times 10^{12} \Msun$. This would be the case if both And XII and
And XIV are not gravitationally bound, which is possible if mechanisms
such as those proposed by \citet{2007MNRAS.379.1475S} are ubiquitous.
It would also require that the proper motion data on M33 and IC10 or
-- perhaps more likely -- the indirectly inferred proper motion of M31
is in error.  Again, such a low estimate for the mass occurs only if
the satellites are moving predominantly radially.

Although it is interesting to ask how low the masses of the Milky Way
and M31 could be, it does produce a mystery in the context of the
Timing Argument, which typically yields larger combined masses.  It is
possible that some of the mass of the Local Group is unassociated with
the large galaxies. Although not the conventional picture, this is
probably not ruled out and there have been suggestions that $\sim
10^{12} \Msun$ may be present in the Local Group in the form of
baryons in the warm-hot intergalactic
medium~\citep{2003Natur.421..719N}. There are few constraints on the
possible existence of dark matter smeared out through the Local Group,
and unassociated with the large galaxies. However, the clustering of
the dwarf galaxies around the Milky Way and M31 does suggest that the
gravity of the dark matter is centered on the prominent galaxies.

\medskip
\noindent
(2) The largest mass estimate we obtained for the Milky Way Galaxy is
$3.9 \pm 0.7 \times 10^{12} \Msun$. This extreme values is driven by
the assumption of tangential anisotropy for the satellites, so that
the measured line of sight velocities also imply substantial
tangential motions as well. The estimate assumes all the satellites
including Leo I to be bound, and the anomalously high proper motion
measurement of Draco to be valid.

Note that the present data allow considerably more scope to increase
the mass of the Milky Way Galaxy than M31. Our largest mass estimate
for M31 is a much more modest $1.6 \pm 0.4 \times 10^{12} \Msun$,
which occurs when we analyse the whole sample incorporating And XII
and And XIV and assume tangentially anisotropic velocity
distributions.

The current concensus is that the two galaxies are of a roughly
similar mass, with M31 probably the slightly more massive of the
two. This though is inferred from indirect properties, such as the
numbers of globular clusters, which correlates with total mass albeit
with scatter, or the amplitude of the inner gas rotation curve. The
stellar halo of M31 is certainly more massive than that of the Milky
Way, although this may not be a good guide to the dark halo. Of
course, it could be that the current concensus is wrong, and that the
Milky Way halo is more massive than that of Andromeda. There is also
some indirect evidence in favour of this -- for example, the typical
sizes of the M31 dwarf spheroidals are large than those of the Milky
Way, which is explicable if the Milky Way halo is denser. However, it
does not seem reasonable to postulate that the mass of the Milky Way
is substantially larger than that of M31. Hence, the very large
estimate of $3.9 \pm 0.7 \times 10^{12} \Msun$ is best understood as a
manifestation of the degeneracy in the problem of mass estimation with
only primarily radial velocity data.

\medskip
\noindent
(3) Our preferred estimates come from accepting Leo I, And XII and And
XIV as bound satellites, whilst discarding the Draco proper motion as
inaccurate.  This gives an estimate for the mass of the Milky Way
within 300~kpc as $2.7 \pm 0.5 \times 10^{12} \Msun$ and for M31 as
$1.5 \pm 0.4 \times 10^{12} \Msun$, assuming the anisotropy implied by
the data ($\beta \approx -4.5$). The error bars are just the
statistical uncertainty and do not incorporate the uncertainty in
anisotropy or sample membership.  In view of this, it is not possible
to decide which of the Milky Way galaxy or M31 is the more massive
based on the kinematic data alone.

These values for the masses are attractive for a number of reasons.
First, the mass ratio between the Milky Way and M31 is of roughly of
order unity, which accords with a number of other lines of
evidence. Second, the values allow most of the dark matter in the
Local Group implied by the Timing Argument to be clustered around the
two most luminous galaxies. Third, they are within the range found for
cosmologically motivated models of the Milky Way and M31
\citep{2008MNRAS.384.1459L}.

We prefer to assume the anisotropy implied by the admittedly scanty
data on the proper motions of the satellites.  However, for
completeness, we quickly sketch the effects of dropping this
assumption.  If the velocity distribution is isotropic, or even
radially anisotropic as suggested by the simulations, then the mass of
the Milky Way becomes $1.4 \pm 0.3 \times 10^{12} \Msun$ or $1.2 \pm
0.3 \times 10^{12} \Msun$ respectively. Similarly for M31, the values
are $1.4 \pm 0.4 \times 10^{12} \Msun$ (isotropy) or $ 1.3 \pm 0.4
\times 10^{12} \Msun$ (radially anisotropic).

\medskip
\noindent
The greatest sources of uncertainty on the masses remain the role of
possibly anomalous satellites like Leo I and the velocity anisotropy
of the satellite populations. There is reason to be optimistic, as the
{\it Gaia} satellite will provide proper motion data on all the dwarf
galaxies that surround the Milky Way and M31, as well as many hundreds
of thousands of halo stars. The analysis that we have carried out here
indicates that proper motions are important if we wish to increase the
accuracy of our estimates, as well as understand the dynamical nature
of objects like Leo I. While we are not yet able to exploit the proper
motions, {\it Gaia} will allow us to do so.

\section*{Acknowledgments} 
NWE thanks Simon White for a number of insightful discussions on the
matter of scale-free estimators. LLW thanks the Science and Technology
Facilities Council of the United Kingdom for a studentship. Work by
JHA is in part supported by the Chinese Academy of Sciences
Fellowships for Young International Scientists (Grant
No.:2009Y2AJ7). JHA also acknowledges support from the Dark Cosmology
Centre funded by the Danish National Research Foundation (Danmarks
Grundforskningsfond). The paper was considerably improved following
the comments of an anonymous referee.


\end{document}